\documentclass[a4paper,11pt]{article}
\pdfoutput=1
\usepackage[T1]{fontenc}
\usepackage{jheppub}
\usepackage[utf8]{inputenc}
\usepackage{float}
\usepackage{hyperref}
\usepackage{graphicx}
\usepackage{rotating}
\usepackage{tikz}
\usepackage{braket}
\usepackage{amssymb}
\usepackage{lmodern}
\usepackage{enumerate}
\usepackage{caption}
\usepackage{subcaption}
\usepackage[nottoc]{tocbibind}
\usepackage{accents}
\usepackage{trimclip}
\usepackage{dutchcal}
\usepackage[titletoc]{appendix}
\usepackage{array}
\makeatletter
\newcommand{\thickhline}{%
    \noalign {\ifnum 0=`}\fi \hrule height 1pt
    \futurelet \reserved@a \@xhline
}
\newcolumntype{"}{@{\hskip\tabcolsep\vrule width 1pt\hskip\tabcolsep}}
\makeatother

\DeclareRobustCommand*{\phat}[1]{{\accentset{(\!\trimbox{0pt 1.1ex}{\ensuremath{\string^}}\!)}{#1}}}

\newcommand{\irow}[1]{
  \begin{smallmatrix}(#1)\end{smallmatrix}
  
  \makeatletter
\newcommand{\thickhline}{%
    \noalign {\ifnum 0=`}\fi \hrule height 1pt
    \futurelet \reserved@a \@xhline
}
\newcolumntype{"}{@{\hskip\tabcolsep\vrule width 1pt\hskip\tabcolsep}}
\makeatother

}

\hypersetup{
    colorlinks=true,
    linkcolor=blue,
    citecolor=green,
    filecolor=magenta,      
    urlcolor=blue,
    linktocpage=true, 
    pdftitle={Overleaf Example},
    pdfpagemode=FullScreen,
    }

\title{First-order phase transitions in Twin Higgs models }
\author{Marcin Badziak}
\author{and Ignacy Nałęcz}
\affiliation{Institute of Theoretical Physics, Faculty of Physics, University of Warsaw, ul. Pasteura 5,\\
PL-02-093 Warsaw, Poland}

\emailAdd{inalecz@fuw.edu.pl}
\emailAdd{mbadziak@fuw.edu.pl}

\abstract{We revisit phase transitions in Twin Higgs (TH) models. We show that strong first-order phase transitions (FOPTs) can occur provided that appropriate source of $\mathbb{Z}_2$ symmetry breaking between the twin and Standard Model (SM) sectors is present. We found FOPTs in two classes of models. First: with hard $\mathbb{Z}_2$ breaking in the scalar potential allowing for FOPT. Second: with $\mathbb{Z}_2$ broken by enhanced Yukawa couplings of twin leptons. We also considered supersymmetric UV completion of the second scenario with light sleptons. The signal of gravitational waves produced during these phase transitions is typically small but can be close to the reach of AEDGE and Einstein Telescope in the case of the FOPT induced by light twin sleptons. Our results open a way to generate SM baryon asymmetry in TH models. }

\begin{document}
\maketitle
\flushbottom

\newpage

\section{Introduction}

Twin Higgs (TH) models~\cite{Chacko:2005pe,Chacko:2005vw,Chacko:2005un} elegantly solve the little hierarchy problem of the SM by introducing a twin (or mirror) copy of the Standard Model (SM). In this framework, the SM-like Higgs boson is a pseudo-Nambu-Goldstone boson (PNGB) of a global SU(4) symmetry and the $\mathbb{Z}_2$ symmetry relating the SM and its twin counterparts allows to avoid breaking of the SU(4) symmetry by quantum corrections. Importantly, the top partners responsible for the naturalness of the electroweak (EW) scale are charged under twin SU(3) gauge group rather than the SM color gauge group, easily avoiding bounds from the LHC. In the UV completions of the TH model utilizing supersymmetry (SUSY)~\cite{Falkowski:2006qq,Chang:2006ra,Craig:2013fga,Katz:2016wtw,badziak2017S,badziak2017M,badziak2017A}, higher dimensions~\cite{Craig:2014aea,Craig:2014roa,Geller:2014kta} or composite Higgs~\cite{Batra:2008jy,Barbieri:2015lqa,Low:2015nqa,Csaki:2015gfd,Contino:2017moj} to solve the big hierarchy problem of the SM, the mass scale of colored top partners can be pushed up to a few-TeV range waiting for the discovery in a future collider.

Even though colored top partners in TH models are expected to be beyond the reach of the LHC, this class of models can be tested at the LHC via the Higgs coupling measurements and searches for additional Higgs bosons~\cite{Barbieri:2005ri,Craig:2015pha,Buttazzo:2015bka,Katz:2016wtw,Ahmed:2017psb}. Even more signatures of TH models are expected in cosmological context due to the richness of the twin sector. Several good dark matter (DM) candidates residing in the twin sector have been identified that can be searched for in direct-detection experiments~\cite{Barbieri:2005ri,Craig:2015pha,GarciaGarcia:2015fol,Farina:2015uea,Craig:2015xla,Barbieri:2016zxn,Barbieri:2017opf,Badziak:2019zys,Terning:2019hgj,Chacko:2021vin,Badziak:2022eag}. 
Mirror sector of TH models may also modify the cosmic microwave background (CMB) and the large-scale structure~\cite{Prilepina:2016rlq,Chacko:2018vss,Bansal:2021dfh}. 
Moreover, the twin sector also generically contributes to dark radiation. As a matter of fact, this contribution is often too big to be consistent with the data and additional mechanism is required to solve this dark radiation problem. Several solutions to this problem have been proposed, including late entropy production that dilutes dark radiation~\cite{Chacko:2016hvu,Craig:2016lyx}, Fraternal Twin Higgs~\cite{Craig:2015pha} where the first two generations of fermions are removed or $\mathbb{Z}_2$ breaking in the Yukawa couplings of light fermions~\cite{Barbieri:2016zxn,Barbieri:2017opf,Harigaya:2019shz}.

In the present article, we investigate phase transitions in TH models and ask a question whether any of such transitions can be first-order. It is well known that first-order phase transition (FOPT) may result in production of gravitational waves (GW)~\cite{Kamionkowski:1993fg,Ellis:2018mja,Caprini:2015zlo,Caprini:2019egz} and may also allow for baryogenesis~\cite{Kuzmin:1985mm,Shaposhnikov:1986jp}, for a review see e.g.~\cite{Cline:2006ts,Morrissey:2012db}. Therefore, a presence of strong FOPT in the cosmological evolution of TH models would provide another handle to test this class of SM extensions. 

The EW phase transition in the SM is known to be smooth~\cite{Arnold:1992rz,Kajantie:1995kf,Kajantie:1996mn,Kajantie:1996qd} but extending the Higgs sector by adding a singlet may result in the first-order EW phase transition, see~e.g.~refs.\cite{Barger:2007im,Brauner:2016fla,Curtin:2014jma}. The Higgs sector of TH models consists of the twin Higgs boson which is a singlet under the SM gauge group. However, the structure of the scalar potential is highly restricted by the approximate SU(4) and $\mathbb{Z}_2$ symmetries so it is not obvious whether such extension allows for the first-order phase transition. 

Phase transitions in TH models have been studied in ref.~\cite{Fujikura:2018duw} reaching the conclusion that they are not first-order. However, the analysis of ref.~\cite{Fujikura:2018duw} assumed that twin EW phase transition, in which twin Higgs field obtained vacuum expectation value (VEV), occurred first and it finished before the start of the EW transition. Such assumption allowed for investigating twin EW and standard EW phase transitions separately and only one field was dynamical at a time. 

In the present paper we show that one-field approximation for phase transitions is not always valid and studying two-field evolution is necessary to get reliable predictions. We also find that strong FOPT can be present if one includes effects of $\mathbb{Z}_2$ symmetry breaking. We demonstrate this in two cases. One: when hard $\mathbb{Z}_2$ breaking is present in the tree-level potential. Two: when $\mathbb{Z}_2$ symmetry is broken in Yukawa couplings of light fermions. 

In the first scenario, we revisited the parameter space of the TH model with an unspecified UV completion (we refer to this scenario as UV agnostic TH) and found a new region where the naturalness can be combined with viable SM Higgs phenomenology, which was not considered in many previous analyzes~\cite{Katz:2016wtw, Fujikura:2018duw}. The FOPT can be present only in this new region which is characterized by hard $\mathbb{Z}_2$ breaking and negative $\mathbb{Z}_2$-preserving but $SU(4)$-breaking quartic coupling. 

In the second scenario, we found FOPTs induced by $\mathbb{Z}_2$ breaking introduced by heavy twin leptons for some range of twin lepton Yukawa couplings. In a supersymmetric UV completion of this scenario, we found that light sleptons can induce even stronger FOPT because they enlarge the barrier through which the tunneling proceeds. For very large values of the Yukawa couplings the EW symmetry is not restored for temperatures up to above 1~TeV. 

The FOPTs we found in all considered scenarios  do not allow for a direct EW baryogenesis in the SM sector while rendering the EW-like baryogenesis in the twin sector possible. Following the lines of darkogenesis models~\cite{Shelton:2010ta}, the excess of twin baryonic matter could be then transfered to the visible sector. The concept of darkogenesis is also motivated by the tight constraints on the electric dipole moment of the electron~\cite{ACME:2018yjb}. Baryogenesis from a dark phase transition has recently been investigated in models with mirror SM gauge group and SM-like particle content in refs.~\cite{Hall:2019rld, Ritter:2021hgu}. Those studies did not aim to solve the hierarchy problem so the TH mechanism was not introduced there and the FOPT was obtained thanks to more freedom in the parameter space allowing, for example, the dark Higgs to be lighter than the SM Higgs. In consequence, the dynamics of those phase transitions is different from that we found in TH models.

We also present numerically simulated spectra of stochastic GW background which can be emitted in the FOPT induced either by hard $\mathbb{Z}_2$-breaking in the scalar potential or by asymmetric Yukawa couplings. In the UV agnostic TH, we found the GW signal much too weak to be detected in the foreseeable future. On the other hand, in supersymmetric extension of TH with light sleptons and enhanced Yukawa couplings of twin leptons the GW signal is much stronger, due to larger barrier created by the thermal correction from sleptons, but may be within reach of the future GW detectors only in a small corner of the parameter space.

The rest of the article is organized as follows. In section \ref{sec:V_eff} we introduce the TH model and its free parameter space, and recall the one-loop effective potential together with leading-order thermal corrections to the scalar potential. Section \ref{sec:pureTH} contains discussion of two possible scenarios of the EW and twin EW symmetry breaking and their dependence on explicit $\mathbb{Z}_2$ symmetry breaking between the SM and twin sector. In section \ref{sec:THnum} we present the numerical evidence for the strong FOPT taking place in SM and twin sectors in UV agnostic TH model with hard $\mathbb{Z}_2$ breaking and identify regions in the parameter space where such transition occurs. Section \ref{sec:lept} is devoted to the analysis of phase transitions in the UV agnostic TH model with $\mathbb{Z}_2$ breaking in lepton Yukawa couplings, whereas in section \ref{sec:SUSY} we consider supersymmetric TH model with light sleptons and large twin lepton Yukawa couplings and analyze their impact on scalar phase transitions. Section \ref{sec:GWbackground} covers the computation of the GW signal emitted in the FOPTs that were found in the preceding sections. Finally, section \ref{sec:summary} contains the discussion and summary of our results, as well as some comments on possible future work.

\section{Twin Higgs scalar potential}\label{sec:V_eff}
\subsection{The tree-level potential}\label{sec:Vtree}
The model contains two scalar doublets: $H_{A}$ which transforms under the SM $SU(2)_L\times U(1)_Y$ gauge group and its twin $H_B$ which transforms under the twin gauge group $SU(2)_{L'}\times U(1)_{Y'}$. The most general tree-level potential of the TH model reads
\begin{equation}\label{eq:VtreeH}
V_{\text{tree}}(H_A,H_B)=\lambda (|H_{A}|^2+ |H_{B}|^2-f_0^2/2)^2+\kappa (|H_{A}|^4+|H_{B}|^4)+f_0^2\sigma |H_{A}|^2+\rho |H_{A}|^4 \,,
\end{equation}
with 
\begin{equation*}
    H_{I}=\begin{pmatrix}\chi^{+}_I
    \\ \frac{1}{\sqrt{2}}(h_{I}+i\,\chi^0_I)\end{pmatrix} \,.
\end{equation*}
Further on we exploit the unitary gauge where $\chi^{+}_I$ and $\chi^{0}_I$ are eaten by longitudinal degrees of freedom of the SM and twin $W^{\pm}$ and $Z^{0}$ vectors. Thus, one can substitute $H_{I}=\left(\begin{matrix}0 \\ \frac{h_{I}}{\sqrt{2}}\end{matrix}\right)$ into \eqref{eq:VtreeH}, obtaining
\begin{equation}\label{eq:V_tree}
V_{\text{tree}}(h_A,h_B)=\frac\lambda 4 (h_{A}^2+ h_{B}^2-f_0^2)^2+\frac \kappa 4 (h_{A}^4+h_{B}^4)+\frac{f_0^2\sigma}{2} h_{A}^2+\frac{\rho}{4} h_{A}^4 \,.
\end{equation}
Here, $\lambda$, $\kappa$, $\sigma$, $\rho$ and $f_0$ are free real parameters. $f_0$ is the scale of the spontaneous $SU(4)$ symmetry breaking and is related to the VEVs of SM and twin Higgs bosons: $f_0^2\approx v_A^2+v_B^2$, where $v_I\equiv\langle h_I \rangle$.
While $f_0$ and $\lambda$ must be positive, there are generically no restriction on the sign of the couplings that explicitly break the $SU(4)$ symmetry~\footnote{In the past literature an arbitrary sign of $\sigma$ was considered while $\kappa$ was assumed to be positive~\cite{Barbieri:2005ri,Beauchesne:2015lva,Katz:2016wtw}. This is well justified as long as one considers hard $\mathbb{Z}_2$ breaking coupling $\rho\leq0$. However, in the general case there is no reason to exclude negative values of $\kappa$ (see for example ref.~\cite{Ahmed:2017psb}).}.
In order to guarantee the approximate $SU(4)$ and $\mathbb{Z}_2$ symmetries which protect the SM Higgs mass parameter from the quadratically divergent loop corrections, the first term in the potential \eqref{eq:V_tree} has to be dominant. Hence, the coupling $\lambda$ should dominate over the other dimensionless scalar couplings: $|\kappa|$, $|\sigma|$, and $|\rho|$.~\footnote{This condition together with conditions for the SM Higgs mass and VEV given in appendix \ref{app:VEVandMass} guarantee that the effective potential is bounded from below.} Nevertheless, the symmetry breaking interactions in the tree-level potential play crucial roles: the $SU(4)$ breaking but $\mathbb{Z}_2$ preserving $\kappa$ provides the non-zero mass of the SM-like Higgs while the softly $\mathbb{Z}_2$-breaking $\sigma$ is necessary to obtain a mild hierarchy $v_B\gtrsim3v_A$ between the VEVs of the SM and twin Higgs bosons, which is required by the LHC Higgs coupling measurements~\cite{Ahmed:2017psb}.
The hard $\mathbb{Z}_2$ breaking interaction controlled by $\rho$ is not absolutely necessary and was often neglected in the past but it can be present in some UV completions of TH models~\cite{Katz:2016wtw}. 

The full expressions for the tree-level VEVs and masses of SM and twin Higgs bosons were given in appendix \ref{app:VEVandMass}. Here, for better clarity, we present the formulas obtained in the limit $\lambda\gg\kappa,\;\sigma,\;\rho$ and $v_A/f_0\ll1$
which is often refereed to as the pseudo Nambu-Goldstone boson approximation. They read
\begin{equation}\label{eq:VEV-s}
v_A^2\approx f_0^2\frac{\kappa-\sigma}{4\kappa+2\rho},\qquad v_B^2\approx f_0^2-v_A^2 \,,
\end{equation}
and
\begin{equation}\label{eq:masses}
m_h^2\approx 2v_A^2(2\kappa+\rho),\qquad m_{h'}^2\approx 2 f_0^2 \lambda,
\end{equation}
where $m_h$ denotes SM-like Higgs boson mass, and $m_{h'}$ corresponds to its twin counterpart.

By setting $v_A=246\;\text{GeV}$ and $m_h=125\;\text{GeV}$ one fixes tree-level values of two out of five free parameters~\footnote{The elimination of free parameters with the one-loop conditions for SM Higgs mass and VEV at $T=0$ is a tedious numerical task. In spite of this, one-loop improved conditions were used for all presented numerical results.}. It is convenient to eliminate $\sigma$ and $\rho$ and work with $\lambda$, $\kappa$ and $f_0$.

The $SU(4)$-breaking scale $f_0$ lays in a very specific range. The consistency with the LHC data on the production and decays of the 125 GeV Higgs requires $f_0\gtrsim3 v_A$~\cite{Ahmed:2017psb}. On the other hand, in TH models tuning grows with $f_0^2/v_A^2$ so to preserve naturalness of the model $f_0$ should not be too large. For example, avoiding tuning worse than 10~\% requires $f_0\lesssim5 v_A$.

\subsection{Loop and thermal corrections}\label{sec:loop}

In order to trace phase transition path, one should compute full effective potential, consisting of zero temperature loop and thermal corrections to the tree-level part \eqref{eq:V_tree} 
\begin{equation}\label{eq:V_tot}
V_{\text{tot}}(h_A,\;h_B)=V_{\text{tree}}(h_A,\;h_B)+V_{\text{therm}}(h_A,\;h_B)+V_{\text{CW}}(h_A,\;h_B).
\end{equation}
Throughout this work we use the Coleman-Weinberg one-loop approximation for zero temperature corrections, computed in the $\overline{\text{MS}}$ renormalization scheme~\cite{Coleman:1973jx}
\begin{equation}\label{eq:VCW}
\begin{aligned}
V_{\text{CW}}(h_A,\;h_B)=\sum_{i\in\text{bosons}}\frac{n_{i} }{64 \pi^2} m^4_{i}(h_A,\;h_B)\left(\log (\frac{m^2_{i}(h_A,\;h_B)}{\mu^2})-c_i\right)\\
-\sum_{i\in\text{fermions}} \frac{n_{i} }{64 \pi^2}m^4_{i}(h_A,\;h_B)\left(\log \frac{m^2_{i}(h_A,\;h_B)}{\mu^2}-\frac{3}{2}\right),
\end{aligned}
\end{equation}
where $n$ is the number of the field degrees of freedom and the coefficients $c_i$ are equal to $5/6$ in case of vectors and $3/2$ in case of scalars. We set the renormalization scale to $\mu=f_0$ and computed the one loop corrections to tree-level values of $\kappa$ and $\sigma$ in order to preserve correct mass and the VEV of the SM Higgs.

The one-loop effective potential \eqref{eq:VCW} and all the observables derived from it depend on the choice of renormalization scale $\mu$. This is unfortunately unavoidable if one works with the leading order approximation of the effective potential.
To estimate the renormalization scale dependence of our results we computed the temperature at which global minimum becomes metastable, called critical temperature $T_c$, using different choices of $\mu$. The critical temperature is the physically meaningful quantity, since at this temperature one expects the onset of the scalar phase transition in the universe. In the FOPT that we discuss in section \ref{sec:pureTH} for $\mu\in[200\text{ GeV},\,2000\text{ GeV}]$ the critical temperature computed using loop and thermally corrected potential \eqref{eq:V_tot} varies by no more than $15\%$. This strongly indicates that any reasonable choice of renormalization scheme leads to roughly the same physical insight. 

Thermal corrections to the scalar potential are given by~\cite{Quiros:1999jp}
\begin{equation}\label{eq:V_therm}
V_{\text{therm}}(h_A,\;h_B)=\frac{T^4}{2\pi^2}\sum_{i\in\text{bosons}}n_i\text{J}_{B}(\frac{m^2_{i}}{T^2})-\frac{T^4}{2\pi^2}\sum_{i\in\text{fermions}}n_i\text{J}_{F}(\frac{m^2_{i}}{T^2})
\end{equation}
where
\begin{align}\label{eq:therm}
J_B(x)=\int_0^{\infty}dk\, k^2 \log[1-\exp(-\sqrt{k^2+x})],\\
J_F(x)=\int_0^{\infty}dk\, k^2 \log[1+\exp(-\sqrt{k^2+x})].
\end{align}
At high temperatures the effective potential receives large contributions from multi-loop bosonic diagrams which diverge in the infrared. This problem is usually solved by the resummation of the leading-order contributions of these diagrams~\cite{Arnold:1992rz}
\begin{equation}
V_{\text{ring}}=-\sum_{i\in\text{bosons}}\frac{n_{i}T}{12 \pi}[(\overline{m}^2_{i})^{\frac{3}{2}}-(m^2_i)^{\frac{3}{2}}],
\end{equation}
or, more exactly, by replacing tree level masses in $V_{\text{therm}}$ and $V_{\text{CW}}$ by their thermal analogues $\overline{m}^2_{i}$~\cite{Parwani:1991gq}. In this work, we adopt the latter procedure. However, even after the resummation we expect our one-loop approximation of the effective potential to be robust if
\begin{equation}\label{eq:CondPert}
\frac{h_I(T)}{T} > g.
\end{equation}
Here, $h_I(T)$ denotes the Higgs VEV in the sector with broken $\text{SU}(2)$ symmetry at temperature $T$ and $g$ is the weak coupling. For more extensive discussion of the resummation and criterion \eqref{eq:CondPert} check appendix \ref{app:resummation}.

\section{Phase dynamics in Twin Higgs models\label{sec:pureTH}}

The classification of possible scenarios for the evolution of fields is an essential step towards finding the regions in parameter space, where the scalar transition leads to events interesting from phenomenological point of view. In general, during thermal evolution of the Universe the expectation values of dynamical scalars follow some specific path in $h_A-h_B$ plane. The scalar phase transition order and dynamics critically depend on its shape. 

We assume that in the considered model at sufficiently high temperatures the $SU(4)$ symmetry is restored, i.e. the global potential minimum is at $h_A=h_B=0$~\footnote{As far as the one-loop approximation of the effective potential is considered, one could show that at sufficiently high temperatures thermal corrections to the scalar masses would dominate field-dependent contribution and force the symmetry restoration~\cite{Kilic:2015joa}.}. Our aim is now to determine which of the fields $h_A$ or $h_B$ acquires its expectation value first. It is a well-known fact that in the non-abelian gauge field theories the perturbative expansion of the effective potential breaks down at high temperatures at which the symmetry is restored~\cite{Arnold:1994bp, Linde:1980ts}. The major problem is that we do not control the radiative corrections from the transverse modes of vector fields. Thus, the perturbative approximation of the effective potential \eqref{eq:V_tot} is, in general, inappropriate tool for analyzing phase transitions at the origin.

Nevertheless, a direction at which the $\text{SU}(4)$ symmetry is broken depends solely on the $\mathbb{Z}_2$-breaking part of the effective potential while the large corrections from heavy vectors are $\mathbb{Z}_2$-symmetric\footnote{This is true only if the SM and TS gauge couplings are identical. Apparently, the TH naturalness requires them to be very close, so with no lost of generality, one can set them formally equal.}. For this reason, in the symmetric phase and its vicinity, one can simply ignore large, unknown contributions from gauge sectors and use the one-loop resummed effective potential \eqref{eq:V_tot} which encodes all the necessary information about the $\mathbb{Z}_2$-breaking in the Lagrangian.

At the temperatures at which symmetry is restored, the effective potential is well approximated by the polynomial in fields (see appendix \ref{app:expansion}). The cubic terms will be tiny, because at high temperatures mass of each particle is dominated by its thermal corrections. Without significant cubic terms, the field cannot tunnel from symmetric phase, and the only way to escape is the continuous phase transition. 

When the continuous transition takes place, the global minimum moves away from the origin. This can happen only if the second order derivative computed at the origin changes sign from positive to negative
\begin{align}
&\frac{\partial^2 V_{\text{tot}}}{\partial h_{A}^2}\bigg|_{h_A=0,\,h_B=0}=\:\zeta_A T^2-\lambda f_0^2(1-\frac{\sigma}{\lambda})\label{eq:mA(0)}\\
&\frac{\partial^2 V_{\text{tot}}}{\partial h_{B}^2}\bigg|_{h_A=0,\,h_B=0}=\:\zeta_B T^2-\lambda f_0^2\label{eq:mB(0)}
\end{align}
where $\zeta_A$ and $\zeta_B$ are thermal mass coefficients which depend on the SM and TS couplings, respectively. The mixed derivative obviously vanishes.

As the Universe cools down, both derivatives decrease and eventually one of them reaches zero. At this point, continuous phase transition occurs, and the minimum leaves the origin and starts moving in $h_A$-$h_B$ plane. Let $T_A$ and $T_B$ denote temperatures at which mass of the SM-like and twin Higgs boson, respectively, would vanish. Then, one may compute the ratio 
\begin{equation}\label{eq:ratio}
\frac{T_A}{T_B}=\sqrt{\frac{\zeta_B}{\zeta_A}}\sqrt{1-\frac{\sigma}{\lambda}}.
\end{equation}

By evaluating the right-hand side of the above equation one finds which of the fields $h_A$ or $h_B$ will first acquire non-zero VEV. The above reasoning was presented in ref.~\cite{Fujikura:2018duw}. However, ref.~\cite{Fujikura:2018duw} considered only the case in which $T_B\gg T_A$. eq.~\eqref{eq:ratio} shows that $T_B> T_A$ if $\sigma>0$ and $\zeta_A\sim\zeta_B$ i.e. neglecting effects of $\mathbb{Z}_2$ breaking in thermal masses. In our analysis we relax the assumption $T_B\gg T_A$ and show that in the presence of $\mathbb{Z}_2$ breaking an analysis of the two-field dynamics is necessary to understand phase transitions in TH models. 

It may seem that the order in which scalar fields acquire expectation values is irrelevant since very soon the VEV of the remaining scalar will also start growing. This, however, seems not to be the case. Once minimum leaves the origin, the equations \eqref{eq:mA(0)} and \eqref{eq:mB(0)} should be corrected (note, that the mixed derivatives are still zero)\footnote{Here, one can safely neglect zero temperature loop corrections, since their contribution is sub-leading close to the temperature of $SU(4)$-breaking transition.}

\begin{align}
&\frac{\partial^2 V_{\text{tot}}}{\partial h_{A}^2}\bigg|_{h_A=0,\,h_B}=\:\tilde{\zeta}_A T^2-\lambda f_0^2(1-\frac{\sigma}{\lambda}-\frac{h_B^2}{f_0^2})\label{eq:mA(phi)},\\
&\frac{\partial^2 V_{\text{tot}}}{\partial h_{B}^2}\bigg|_{h_A,\,h_B=0}=\:\tilde{\zeta}_B T^2-\lambda f_0^2(1-\frac{h_A^2}{f_0^2}).\label{eq:mB(phi)}
\end{align}
From the above equations it follows that non-zero VEV of one field stabilizes the other one by giving positive contribution to its effective mass. Hence, close to the origin the yet unbroken EW or twin EW symmetry can be broken only by the first-order phase transition. On the other hand, when the difference $|h_A-h_B|$ becomes large, thermal loop contributions from transverse modes are not $\mathbb{Z}_2$-symmetric anymore. Thus, for big scalar expectation values the above equation does not hold, and continuous phase transition is possible. In either case, critical temperature, transition order and distance between degenerate minima strongly depend on the sector in which the $\text{SU}(2)$ symmetry is broken first. This in turn is controlled by the terms breaking the discrete $\mathbb{Z}_2$ symmetry between the SM and twin sectors.

Let us first discuss scenario in which the only source of $\mathbb{Z}_2$-breaking are the scalar couplings $\sigma$ and $\rho$. In such a case, one may compute the tree-level $\mathbb{Z}_2$-asymmetric contribution to the thermal masses of the Higgs bosons by evaluating the second order derivative of the thermal potential $\eqref{eq:V_therm}$ at the origin. This way, one obtains $\zeta_A\approx\zeta_B+\rho/2$, which, substituted to equation \eqref{eq:ratio}, yields
\begin{equation}\label{eq:tempratiors}
    \frac{T_A}{T_B}\approx \sqrt{(1-\frac{\sigma}{\lambda})(1-\frac{\rho}{2\,\zeta_A})}.
\end{equation}
Thus, initial direction of symmetry breaking depends on two ratios: 
$\sigma/\lambda$ and $\rho/(2\zeta_A)$. $\rho/(2\zeta_A)$ is usually smaller since the tree-level thermal correction coefficient to the SM Higgs mass $\zeta_A$ is greater than $\lambda$. Hence, it is mainly the sign of $\sigma$ that determines whether $T_A$ or $T_B$ is larger (through the first factor on the rhs of eq.~\eqref{eq:tempratiors}) so for 
$\sigma<0$, $h_A$ first acquires non-zero VEV~\footnote{By definition, $\zeta_A$ is the coefficient of $T^2$ in the SM Higgs thermal mass correction $\Pi_{h_A}(T)$. Looking at the expression for $\Pi_{h_A}(T)$ given in the appendix \ref{app:resummation} one can verify that for major part of considered parameter space $\lambda<\zeta_A$. Still, for a small range of negative $\sigma$, for which $|\rho|/(2\zeta_A)\gtrsim|\sigma|/\lambda$, $h_B$ first acquires non-zero VEV.}.

The twin symmetry between the model sectors can be also broken by couplings which do not enter the tree-level scalar potential \eqref{eq:V_tree}. Any significant $\mathbb{Z}_2$-breaking interactions would generate hierarchy between $\zeta_A$ and $\zeta_B$, which could overpower the effect of $\sqrt{1-\sigma/\lambda}$ factor in eq.~\eqref{eq:ratio}. In fact, by adjusting $\zeta_A$ to $\zeta_B$ ratio, one gains control over the phase evolution onset, irrespectively of $\mathbb{Z}_2$-breaking parameter values in the scalar potential. On the other hand, strong $\mathbb{Z}_2$ breaking in the non-scalar sector would generically induce large radiative corrections to the SM Higgs mass parameter. This will obviously reintroduce fine-tuning to the model unless one considers a UV extension where those radiative corrections are naturally cancelled (an example of such extension is introduced in section~\ref{sec:SUSY}).

In the following sections we describe the results of our numerical analysis. We classify which paths of the thermal evolution of the fields may be present depending on the parameters with a special emphasis of the paths in which strong first-order phase transitions are present. 

\section{$\mathbb{Z}_2$ breaking in the Higgs potential only}\label{sec:THnum}

\begin{table}[t]
\centering
\begin{tabular}{|c|c|c|c|c|c|}
\hline
Benchmark point & $f_0$ & $\lambda$ & $\kappa$ & $\sigma_{\text{tree}}+(\sigma_{\text{loop}})$  & $\rho_{\text{tree}}+(\rho_{\text{loop}})$  \\ \hline
P1 & $4\,v_A$ & $1$ & $-0.1$ & $-0.12+(0.03)$  & $0.35+(-0.07)$ \\ \hline
P2 & $4\,v_A$ & $1$ & $0.1$ & $0.083+(0.024)$  & $-0.054+(-0.072)$ \\ \hline
\end{tabular}
\caption{Benchmark points in TH model with $\mathbb{Z}_2$ symmetry broken only in the scalar potential. The loop corrections to the tree level $\sigma$ and $\rho$ values are given in brackets.}
\label{tab:benchmark0}
\end{table}

\subsection{Paths of thermal evolution \label{ssec:sg0}}
\begin{figure}[t]
    \centering
    \subcaptionbox{Phase diagram for benchmark point P1.
    \label{fig:sigmag0}}{\includegraphics[scale=0.45]{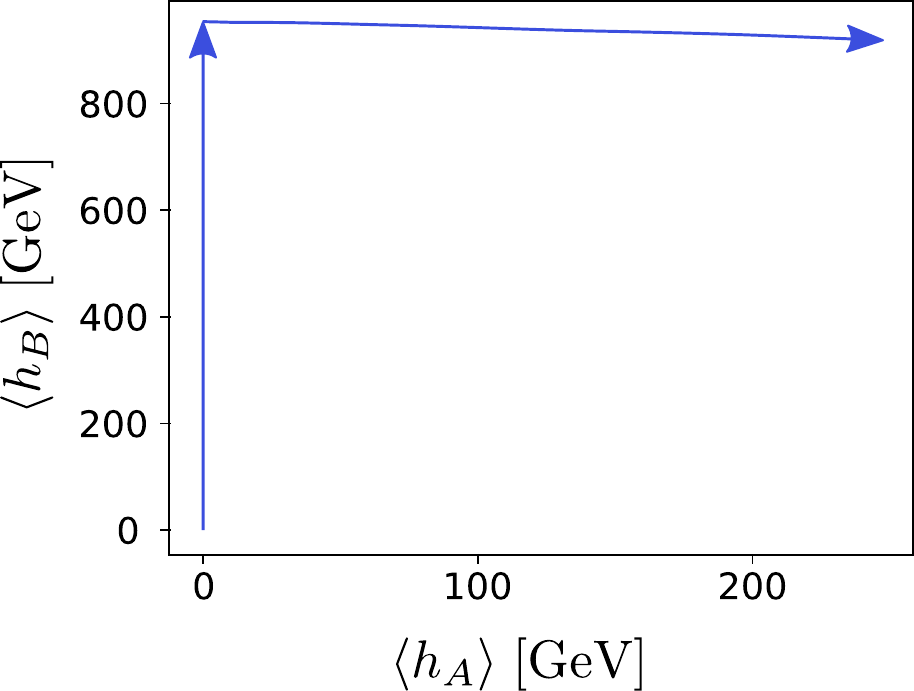}}\hspace{5mm}
    \subcaptionbox{Phase diagram for benchmark point P2. 
    \label{fig:sigmal0}}{
    \includegraphics[scale=0.45]{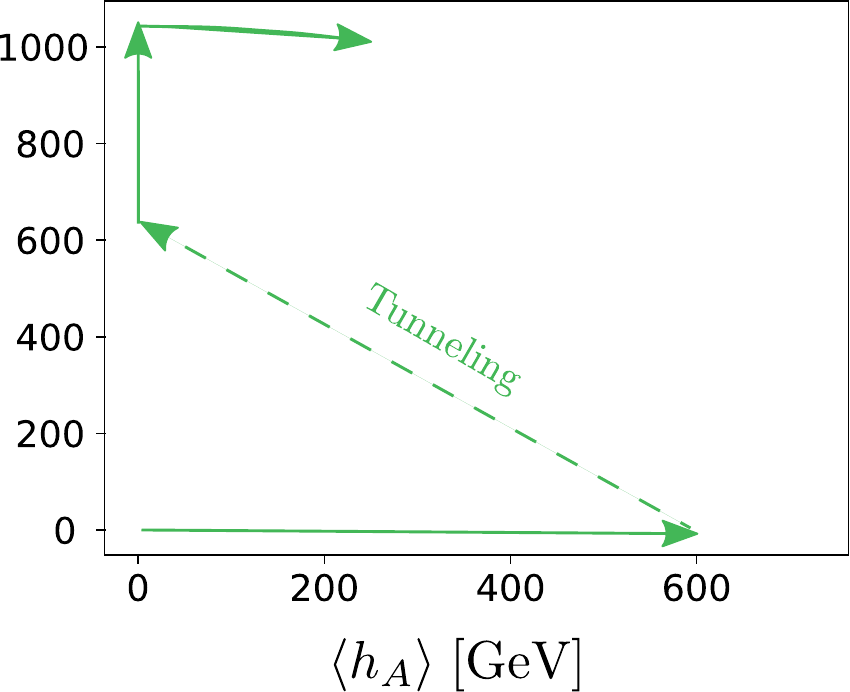}}
    \caption{Evolution of the global minimum of the effective potential \eqref{eq:V_tot} for the two benchmark points in table~\ref{tab:benchmark0} with the opposite signs of $\sigma$. On the right panel dashed arrow connects two degenerate minima, separated by the potential barrier.}
    \label{fig:sigma}
\end{figure}

We start with a scenario in which $\mathbb{Z}_2$ breaking is obtained only via the soft and hard breaking parameters in the tree-level TH potential. In this class of models, two types of paths for the field evolution can be present, which we discuss using two representative benchmark points listed in table~\ref{tab:benchmark0}. We have numerically traced the thermal evolution of a global minimum of the one-loop resummed effective potential \eqref{eq:V_tot}. The resulting paths in $h_A-h_B$ space together with the information about all the phase transitions is shown in figure~\ref{fig:sigma}. As expected from the analytic discussion in the previous section, we obtained different trajectories for the points with positive and negative $\sigma$.

For $\sigma>0$ we found that both the path shape and phase transition order are consistent with predictions made on analytical grounds in ref.~\cite{Fujikura:2018duw}. In this regime, the symmetry is initially broken in the $h_B$ direction and scalar phase does not undergo any tunneling -- all phase transitions were continuous, see figure~\ref{fig:sigmag0}.

On the other hand, for non-negligible negative values of $\sigma$~\footnote{Note that the benchmark point with $\sigma<0$ is characterised by $\rho>0$ and $\kappa<0$ which is a consequence of the constraints on the parameter space from the minimization conditions and the measured Higgs mass of 125 GeV.} we found a new type of paths, which was not yet described in the literature in context of TH models and is shown in figure~\ref{fig:sigmal0}. Here, the symmetry breaking starts in the $h_A$ direction and then, as the universe cools down, the $h_A$ accumulates its expectation value while $h_B$ remains zero. We found that at some point global minimum exchanges with the local one, separated from it by the potential barrier. Such behavior strongly indicates tunneling from the local to the global minimum. During this phase transition $h_A$ moves to zero and the EW symmetry is restored while $h_B$ gets a VEV and breaks twin EW symmetry. Eventually, due to high twin Higgs VEV, masses of strongly coupled twin particles become larger than the temperature of the Universe and decouple from the thermal bath. At this point one recovers smooth one-dimensional dynamics with $h_A$ behaving as Higgs field in unmodified SM and $h_B$ fixed by the effective-field-theory (EFT) approximation
\begin{equation}\label{eq:EFT}
   h_B\approx\sqrt{f_0^2-h_A^2}.
\end{equation} 
The scalar phase evolution proceeds in this manner until both, $h_A$ and $h_B$, roll down to their zero temperature VEVs.

The parameters characterizing both phase transitions are given in table~\ref{tab:benchmark1}. To ensure that in the case with $\sigma<0$ the FOPT really takes place, we have numerically found the full instanton solution for the profile of the bubble of the new phase (thus providing a simple existence prove). The aforementioned solution was obtained with the so-called path deformation algorithm implemented in the \verb|cosmoTransitions| package~\cite{Wainwright:2011kj}. 

Finally, let us comment on the accuracy of our one-loop computations. At both benchmark points the condition \eqref{eq:CondPert} was not satisfied in the vicinity of the phase transition. Therefore, we expect that for chosen points higher loop terms in the effective potential were non-negligible at $T\sim T_c$ and thus the one-loop results are merely qualitative estimates of the ongoing processes. As we will show in the next section, for different parameter values one can obtain larger $h_I(T_c)/T_c$ which renders our results more reliable.

\begin{table}[t]
\centering
\begin{tabular}{|c|c|c|c|c|c|c|c|}
\hline
$\sigma$ & tr. type & $T_c\;\text{[GeV]}$ & $T_n\;\text{[GeV]}$ & $v_A^{\text{ini}}\;\text{[GeV]}$ & $v_B^{\text{ini}}\;\text{[GeV]}$ & $v_A^{\text{fin}}\;\text{[GeV]}$ & $v_B^{\text{fin}}\;\text{[GeV]}$  \\ \hline
$0.11$     & continuous       &    $183$   &    --   &         $0$           &        $946$          &          $0$         &        $946$         \\ \hline
$-0.09$    & $1$-st order       &   $1235$    &   $1172$    &       $0$       &        $652$         &        $704$     &       $0$          \\ \hline
\end{tabular}
\caption{Parameters characterizing phase transition obtained for benchmark points from table~\ref{tab:benchmark0} (for visualization see diagrams in figure~\ref{fig:sigma}). For the case with negative $\sigma$ we have computed the full instanton solution and used it to obtain the temperature at which the bubbles of the new phase start nucleating which we refer to as nucleation temperature $T_n$. As one can see, it only slightly differs from the critical temperature. $v_I^{\text{ini}}$ ($v_I^{\text{fin}}$) corresponds to the VEV of $h_I$ at the beginning (end) of the phase transition.}
\label{tab:benchmark1}
\end{table}

\subsection{Systematic scan over the parameter space}

As we have shown in the previous section, at some benchmark points the scalar fields would likely undergo tunneling between the distant potential minima. Such a process could have significant phenomenological consequences. If the FOPT turns out to be sufficiently strong, it may open a way for EW-like baryogenesis and/or generation of stochastic gravitational waves background within TH. We are now going to analyze the model parameter space, classifying the scalar phase transition order and its strength both in the SM and twin sectors of the theory. Hereafter, we employ the standard condition for the strong transition which originates from the requirement that needs to be satisfied to decouple SM EW sphalerons\footnote{One may wonder if the standard condition for sphaleron decoupling derived for the SM-like models is also applicable to TH. We have checked, that due to structure similarities between SM and TH, this condition should be roughly the same for both theories.}
\begin{equation}\label{eq:CondvT1}
    \frac{\Delta v_I (T_{n})}{T_{n}}>1,
\end{equation}
where $\Delta v_I$ denotes the difference between the Higgs $I$ VEVs at the \textit{end} and at the \textit{beginning} of the tunneling while $T_{n}$ is the temperature at which stable bubbles of the new phase nucleate. We found that in the context of TH the temperature of bubble nucleation is always well approximated with the critical temperature $T_c$ (see for example table~\ref{tab:benchmark0}). Since computation of the latter is much faster, in all the scans we use $\Delta v_I (T_{c})/T_{c}$ rather than $\Delta v_I (T_{n})/T_{n}$.

If one allows for the modifications of the standard cosmological scenario by allowing for the new relativistic species active at high temperatures, the Hubble expansion rate during phase transition could be significantly boosted. This weakens the sphaleron decoupling condition, down to~\cite{Beniwal:2017eik} 
\begin{equation}\label{eq:CondvT2}
    \frac{\Delta v_I (T_c)}{T_{c}}>0.5.
\end{equation}

The ratio $\Delta v_I (T_{c})/T_{c}$ used to determine the EW sphalerons activity at phase transition encodes also an information about the quality of the perturbative approximation (see the condition \eqref{eq:CondPert}). The larger is this ratio in both sectors, the more robust are results obtained with the one-loop effective potential. Thus, $\Delta v_I (T_{c})/T_{c}$ quantifies both the strength and reliability of the phase transition and is the natural parameter to be computed in the scans over the model parameter space.

It turns out that the condition for sphaleron decoupling in modified cosmological scenarios \eqref{eq:CondvT2} is slightly weaker than the perturbativity condition \eqref{eq:CondPert}. Nevertheless, rough character of the latter leaves the possibility that the results with $0.5<\Delta v_I (T_c)/T_{c}\lesssim g$ may still lead to approximately correct results. Since we would like to be as general as it is possible, we are going to confront our results with both the ordinary and weak sphaleron decoupling conditions.

In order to check where strong and weak conditions for FOPT are met, we run the scan over the whole parameter space. Our program traced global minimum of the effective potential \eqref{eq:V_tot} for approximately $2\, 500$ points in the parameter space. The minimum tracing was handled with the help of dedicated numerical package \verb|CosmoTransitions| which provides an accurate and fast solution for this task~\cite{Wainwright:2011kj}. 

 \begin{figure}[t]
    \centering
    \subcaptionbox{Transition strength in the SM sector}{\includegraphics[scale=0.45]{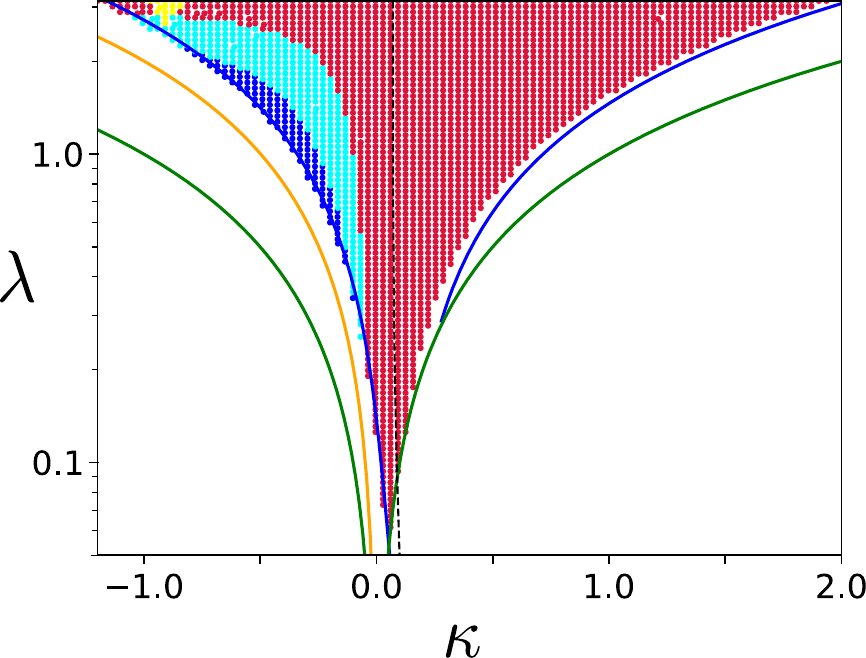}}\hspace{5mm}
    \subcaptionbox{Transition strength in the TS sector}{\includegraphics[scale=0.45]{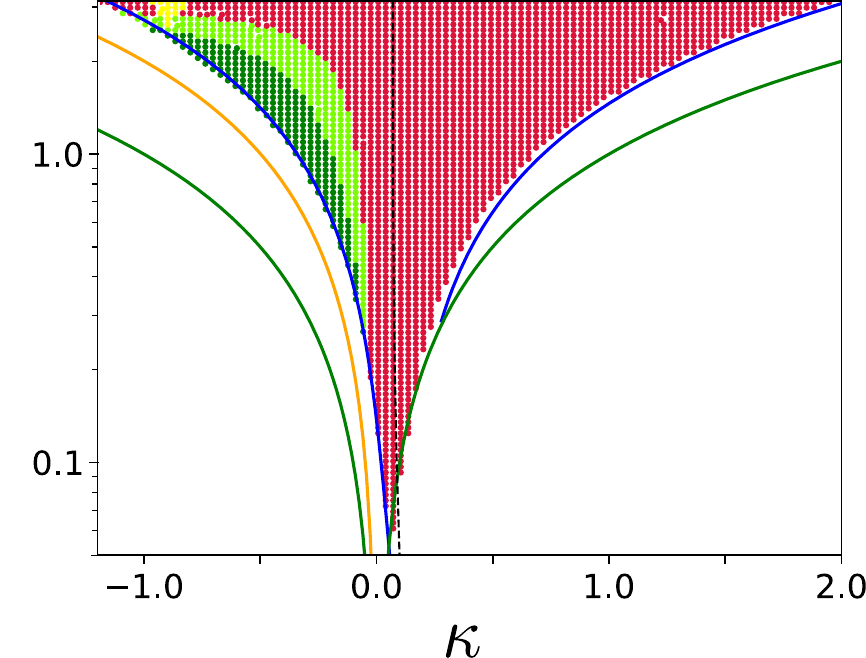}}
        \caption{The strength of the phase transition in the TH model with $\mathbb{Z}_2$ broken only in the Higgs potential in $\kappa$-$\lambda$ plane for $f_0=4\,v_A$. The results for the SM (twin) Higgs are presented on the left (right) panel. The regions where the SM (twin) Higgs transition satisfies \eqref{eq:CondvT1} is marked with dark blue (green) while the regions where transition fulfills \eqref{eq:CondvT2} is plotted with cyan (lime). In the red region the transition was found very weak first-order or continuous. The yellow region is excluded because the $125\; \text{GeV}$ scalar corresponds to twin instead of SM Higgs. The region above the green, blue and orange curves satisfy the corresponding bounds $|\kappa|<\lambda$, $|\rho|<\lambda$, and $|\sigma|<\lambda$. Black, vartical, dashed line corresponds to $\rho=0$.}
    \label{fig:scan}
\end{figure}

 The results of our computations are plotted in figure~\ref{fig:scan}. It follows from the plots that there exists a small region in the parameter space, where ordinary sphaleron condition~\eqref{eq:CondvT1} for twin baryogenesis is satisfied~\footnote{Note that sphaleron decoupling conditions for the SM EW baryogenesis are never fulfilled. The ratio $v(T_c)/T_c$ was computed using \textit{initial} value of the SM Higgs VEV since the final value is obviously close to zero. The ratio obtained this way can only be used to roughly estimate the quality of perturbative approximation.}. This region is described by $\kappa<0$ which follows from the fact that first-order phase transition in this scenario requires $\sigma<0$ and hence relatively large positive $\rho$ which, in turn, requires $\kappa<0$ to accommodate the 125 GeV Higgs mass. Even though non-equilibrium dynamics takes place in both, the twin and SM sectors, in the latter EW sphalerons do not freeze-out in the new phase and therefore the formation of baryon asymmetry is not possible. Nevertheless, the obtained results prove that phase transition in TH models can be strongly first-order.

 To ensure that observed strong transitions are not an artifact of our low-order perturbative expansion, we recomputed the ratio $v(T_c)/T_c$, in the model with all gauge couplings set to zero. This eliminates the contributions from gauge bosons which are mainly responsible for bad convergence at $T\sim T_c$. Removing them renders theory far more stable at phase transition. Even after removing gauge boson contributions, which generically boost the transition order, at considerable part of the parameter space FOPT remains sufficiently strong to fulfill the condition \eqref{eq:CondvT1}. This proves that, at least for some parameter values, FOPTs observed in TH do not require potentially unreliable gauge boson contributions and are therefore robust.

We should, however, emphasize that the region where FOPT is strong is characterized by relatively large values of $\rho$. For example, for $\lambda=1$ we found that phase transition in the twin sector fulfills the ordinary sphaleron decoupling condition \eqref{eq:CondvT1} for $\rho\gtrsim0.55$. Even though this satisfies $\rho<\lambda$ so the 125 GeV can be described as PNGB to a good approximation, some amount of tuning between $\kappa$ and $\rho$ is required to obtain the measured Higgs mass for such large values of $\rho$. This is because the 125 GeV Higgs mass implies $2\kappa+\rho\approx0.1$ so values of $\rho\gg0.1$ require additional tuning. One cannot precisely quantify this tuning without specifying UV completion of the model but one should keep in mind that generically there will be some upper bound $\rho$ stemming from naturalness. The weaker twin sphaleron condition \eqref{eq:CondvT2} for $\lambda=1$ leads to $\rho\gtrsim0.27$ for which the tuning in the Higgs mass is less severe.

In conclusion, our results indicate that strong first-order transition, as defined in \eqref{eq:CondvT1}, can take place in the ordinary TH model but may require some tuning of the Higgs mass, whereas the weaker FOPT satisfying \eqref{eq:CondvT2} can be combined with substantial reduction of the tuning of the EW scale by the TH mechanism. 

\section{Explicit $\mathbb{Z}_2$ breaking in lepton Yukawa couplings\label{sec:lept}}

In the previous section we assumed exact mirror $\mathbb{Z}_2$ symmetry between the SM and TS couplings, broken only by the scalar interactions. However, to alleviate the little hierarchy problem, it is sufficient to make TS gauge couplings and the top Yukawa roughly equal to their SM analogues. To avoid large radiative corrections to the twin Higgs mass parameter that would result in fine-tuning one should also impose $\hat{y}_{q}\ll y_t$ for $q\in\{\hat{u},\hat{d},\hat{s},\hat{c},\hat{b}\}$ (here, we use the hat to denote TS fermions and their couplings). At first glance, it may seem that to preserve the naturalness twin lepton Yukawas should also obey $\hat{y}_{l}\ll y_t$ ~\cite{Craig:2015pha}. However, if the cut-off scale for the lepton sector is sufficiently low, which may still be allowed by the LHC searches~\cite{particle2022} for lepton partners, large twin lepton Yukawa couplings may not lead to conflict with the naturalness. Thus, in the present section we investigate the effect of large twin Yukawa couplings on phase transitions in TH models. 
In section~\ref{sec:SUSY} we will consider the case of large twin lepton Yukawa couplings in the framework of SUSY TH model where it does not lead to fine-tuning provided that sleptons are light enough~\cite{Badziak:2019zys}.

The possibility of the mirror symmetry breaking is often exploited in the model variants that aim to resolve the problem of excessive dark radiation which generically appears in TH models, see~e.g.~\cite{Chacko:2016hvu}. The most explicit example is the Fraternal TH in which TS contains only third generation fermions~\cite{Craig:2015pha}. Another solution to the dark radiation problem is to increase twin lepton and quark Yukawa couplings in order to shift the dark QCD phase transition above the TS decoupling temperature~\cite{Barbieri:2016zxn,Barbieri:2017opf}.

To get a better understanding of how the discrete twin symmetry breaking in lepton Yukawa(s) could affect the type and strength of phase transition, one should look again at equation \eqref{eq:ratio}. Let us recall that in TH models strong FOPT is feasible only when $T_A>T_B$. Previously, we considered the specific case with $\zeta_A\sim\zeta_B+\rho/2$ and thus, strong FOPT required $\sigma<0$. Negative $\sigma$ is always obtained at the cost of introducing non-negligible hard $\mathbb{Z}_2$ breaking $\rho$. As it is usually problematic to find UV completion of the TH model which naturally explains big values of $\rho$ (see however~\cite{Katz:2016wtw}), here we consider an alternative approach that makes FOPT possible. It is based on manipulating twin lepton Yukawa couplings in a way that introduces hierarchy between $\zeta_A$ and $\zeta_B$. Since the SM Yukawa couplings for leptons are negligible, the question really is whether introducing twin leptons with large Yukawa couplings would boost the coefficient $\zeta_B$.

To address this question one can compute leptonic contributions to the twin Higgs thermal mass $\zeta_B T^2$. In general, fermions can affect $\zeta_B$ only through the thermal part of the effective potential \eqref{eq:V_therm}. At high temperatures, one can expand fermion thermal function $J_\text{F}(\hat{m}_l^2/T^2)$ (see Appendix~\ref{app:expansion}), to find the correction from lepton species $i$ to the Higgs mass
\begin{equation}
  \Delta \hat{m}^2_{i\rightarrow h_B}=\frac{\partial^2}{\partial h_B^2}\left[-\frac{\hat{n}_i T^4}{2\pi^2}J_F(\frac{\hat{m}_i ^2}{T^2})\right]=T^2\frac{\hat{n}_i \hat{y}^2_{i} }{48}+\mathcal{O}\Big(\big(\frac{\hat{m}^2_i}{T^2}\big)^{\frac{3}{2}}\Big)
   \end{equation}
where $\hat{n}_i$ is the number of degrees of freedom coupled with $\hat{y}^2_{i}$. As one can see, twin leptons provide positive contribution to $\zeta_B$, proportional to their Yukawa couplings squared and the number of degrees of freedom. Whenever lepton couplings (or their number) are sufficiently large, the symmetry is broken in the $h_A$ direction first, so the \textit{necessary} condition for FOPT can be satisfied even without introducing $\rho\neq0$.

At lower temperatures, the impact of twin fermions on TH phase transitions can be understood with the help of EFT approximation \eqref{eq:EFT}. As long as high-temperature expansion of the thermal function $J_\text{F}$ holds, the leading order fermion corrections to the effective potential are of the form
\begin{equation}
\begin{aligned}
  &\delta V_{\text{eff}}= \left.\delta V_{\text{eff}}\right|_{T=0}+T^4\sum_{i\in\text{Fermions}} \Big[\frac{n_i}{24\, T^2}(m^2_i+\hat{m}^2_i)+\mathcal{O}\Big(\big(\frac{m^2_i}{T^2}\big)^{\frac{3}{2}}\Big)\Big]\approx\\
    &\left.\delta V_{\text{eff}}\right|_{T=0}+T^2\sum_{i\in\text{Fermions}} \Big[\frac{n_i}{48} (y^2_i-\hat{y}^2_i) h_A^2+f_0^2\hat{y}^2_i n_i\Big].
    \end{aligned}
\end{equation}

Thus, twin fermions with large Yukawa couplings will oppose EW symmetry restoration. Usually, their effect becomes dominant at temperatures larger than temperature of EW symmetry restoration~$\sim100\;\text{GeV}$, and hence, they break it again pushing global minimum from $h_A=0$, $h_B=f_0$ towards the phase with $h_A=f_0$ and $h_B=0$. 

The impact of heavy fermions on the phase transition in the models where the SM Higgs is PNGB was noticed in~\cite{Matsedonskyi:2020mlz}. 
It was shown that for sufficiently large number of strongly coupled twin fermions the EW symmetry is not restored and the sphaleron decoupling condition \eqref{eq:CondvT1} holds until $T\sim f_0$. However, the number of fermions necessary to get symmetry non-restoration together with sphaleron freeze-out exceeds the number of twin leptons, even after including twin neutrinos\footnote{To get the EW symmetry non-restoration within TH it is necessary to introduce twin leptons \textit{and} quarks with order one Yukawas. Nonetheless, this is possible but requires very specific UV completion to make it consistent with experimental bounds and keep the fine-tuning under control~\cite{Matsedonskyi:2020kuy}.}. Hence, in this section we do not expect symmetry non-restoration.

Let us consider two cases. 
First: with the twin tau Yukawa coupling strongly enhanced and negligible Yukawa couplings of lighter twin leptons. Second: with all twin lepton Yukawas being large, which makes them significantly heavier than their SM counterparts. We assume, for simplicity, that all twin lepton Yukawa couplings are equal in this case.

\begin{figure}[t]
    \centering
    \subcaptionbox{The SM transition strength. \label{fig:scanYSM}}{\includegraphics[scale=0.4]{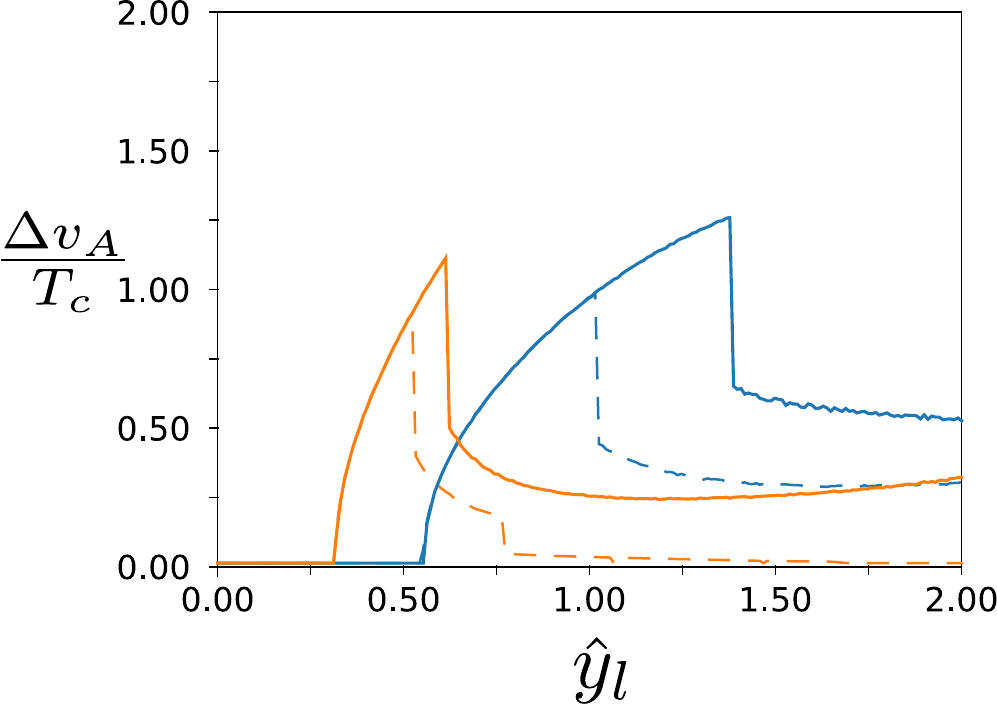}}\hspace{5mm}
    \subcaptionbox{The TS transition strength.\label{fig:scanYDS}}{
    \includegraphics[scale=0.4]{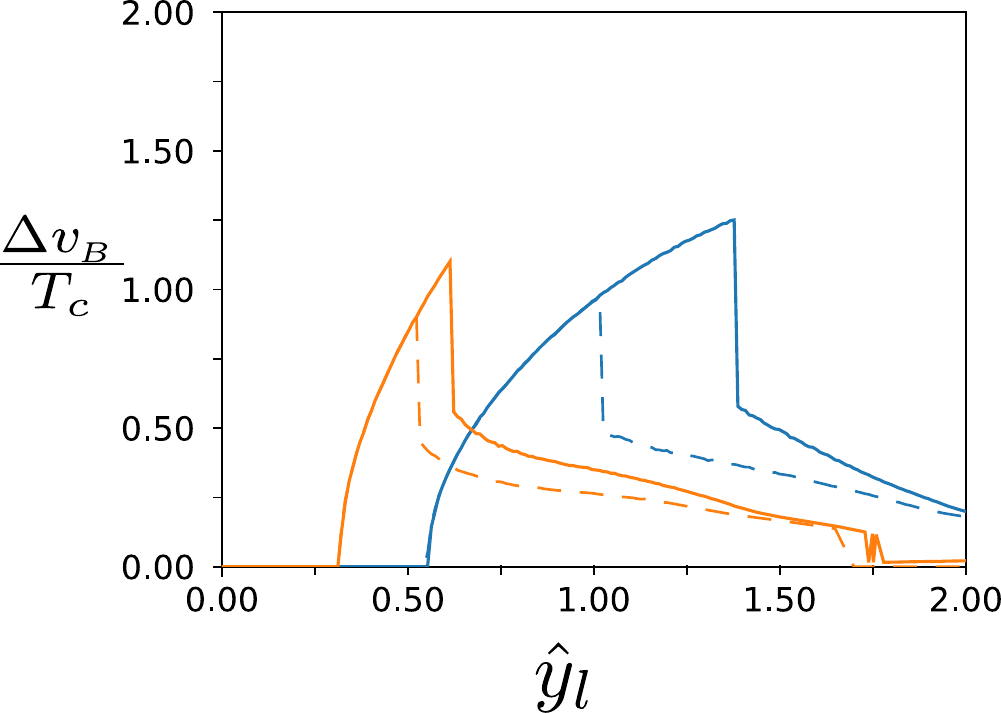}}
    \caption{The phase transition strength $\Delta v_I(T_c)/T_c$ as a function of twin lepton Yukawa coupling $\hat{y}_l$ for $\lambda=1$ and $\rho=0$. Blue (orange) lines correspond to the case with the Yukawa couplings of one (all three) twin leptons equal to $\hat{y}_l$. Dashed and solid lines correspond to $f_0=3v_A$ and $f_0=4v_A$, respectively. }
    \label{fig:scanY}
\end{figure}

Let us start with a discussion of the phase transition strength dependence on twin lepton Yukawa. We focus on the benchmark points with $\lambda=1$ and $\rho=0$ and two different values of the $SU(4)$ breaking scales: $f_0=3\,v_A$ and $f_0=4\,v_A$. The phase transition path is similar to the one presented in figure~\ref{fig:sigmal0}. In figure~\ref{fig:scanY} we present the plot of $\Delta v_I(T_c)/{T_c}$ against the enhanced twin lepton Yukawa coupling. We see that the weak twin sphaleron decoupling condition $v_B(T_c)/T_c>0.5$ is satisfied for $\hat{y}_l$ at least about 0.4 (0.7) if all three (one) twin lepton Yukawa couplings are large. We also see that in each case there is some maximal value of the twin tau Yukawa coupling, $(\hat{y}_l)_c$, above which the strength of the phase transitions decreases with $\hat{y}_l$. Whether the standard twin sphaleron decoupling condition $v_B(T_c)/T_c>1$ is satisfied depends on $f_0$. This condition is satisfied for $f_0=4v_A$ and $\hat{y}_l$ at least about 0.7 (1.1) if all three (one) twin lepton Yukawa coupling is large.

The sharp drop of the phase transition strength at some value of $(\hat{y}_{l})_c$ is caused by the intermediate phase with $(h_A\neq0, h_B\neq0)$ that, for a short time, becomes a global minimum. This intermediate phase is in fact the potential well that arises in the middle of the barrier between $(h_A\neq 0, h_B=0)$ and $(h_A=0, h_B\neq0)$ minima, making it effectively M-shaped. We have numerically verified that the M-shape barrier appears only in the close vicinity of $(\hat{y}_{l})_c$ and above, while for smaller $\hat{y}_{l}$ it is more $\Lambda$-shaped. Therefore, for $\hat{y}_{l}<(\hat{y}_{l})_c$ there is no risk that tunneling will proceed through the intermediate local minimum. For smaller values of $f_0$, $(\hat{y}_{l})_c$ is smaller, as clearly seen in figure~\ref{fig:scanY}.

\begin{figure}[t]
    \centering
    \includegraphics[scale=0.45]{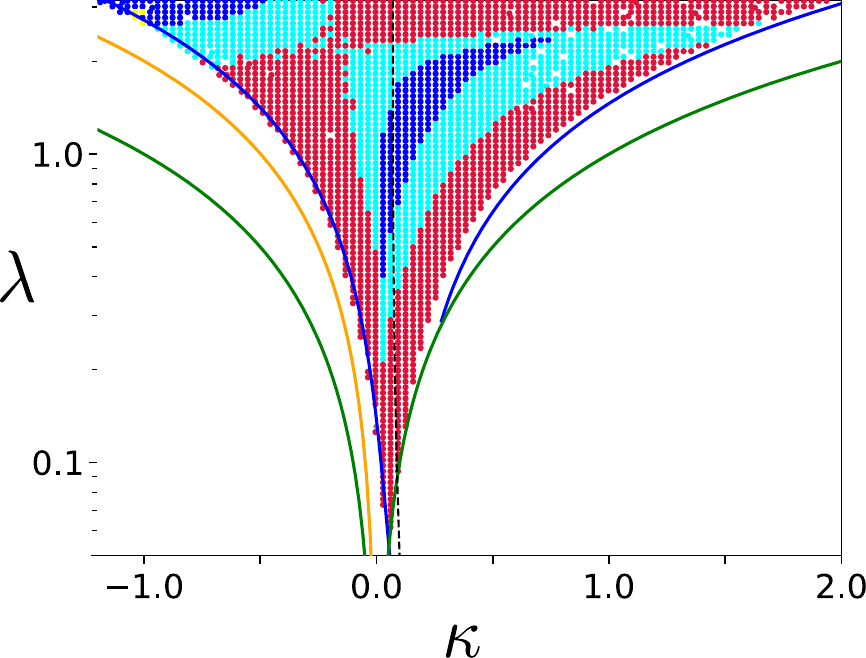}\hspace{5mm}
    \includegraphics[scale=0.45]{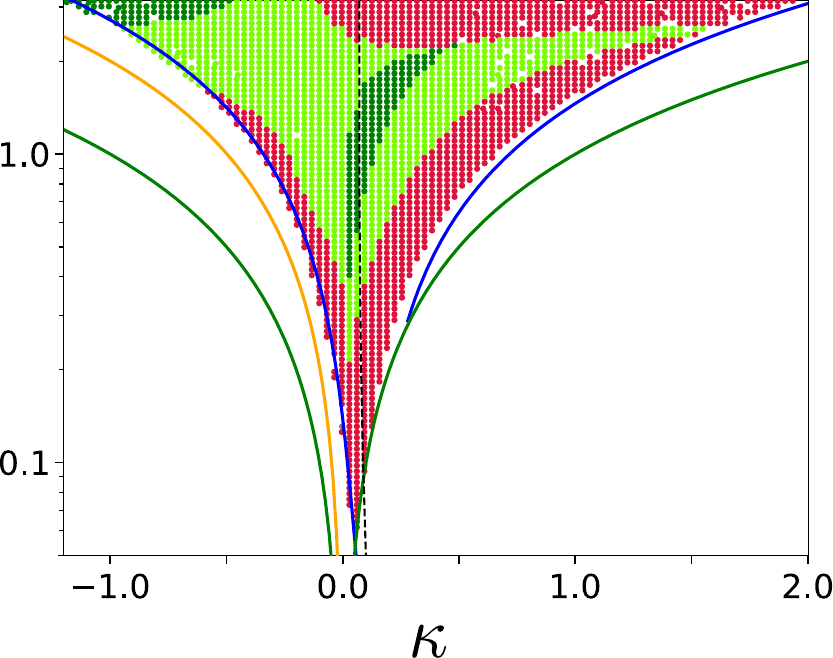}
        \caption{The same as in figure~\ref{fig:scan} but for the case of $\mathbb{Z}_2$ breaking by one twin Yukawa coupling $\hat{y}_{\tau}= 1.34$ corresponding to the maximal strength of the phase transition for the benchmark point with $\lambda=1$, $\rho=0$ and $f_0=4v_A$, as demonstrated in figure~\ref{fig:scanY}.}
    \label{fig:scanlept}
    \end{figure}

To visualize the combined effect of $\mathbb{Z}_2$-breaking in the TH scalar potential and asymmetric Yukawa couplings we performed a scan of parameters $\lambda$ and $\kappa$ which includes also the case of $\rho\neq0$. In the scan we considered the case in which only one twin lepton Yukawa coupling is non-negligible and fixed to $\hat{y}_{l}= 1.34$, which corresponds to the maximal strength of the phase transition for $f_0=4v_A$, according to figure~\ref{fig:scanY}. It is clearly seen in figure~\ref{fig:scanlept} that in such a case much larger parts of parameters lead to FOPT, as compared to the case with no $\mathbb{Z}_2$ breaking in Yukawa couplings, cf.~figure~\ref{fig:scan}. Moreover, the region of parameters in which the strong twin sphaleron decoupling condition is satisfied is now characterized by positive values of $\kappa$ while $\rho$ can have both signs or vanish. For $\rho=0$ strong FOPT corresponds to values of $\lambda$ in the range between about $0.6$ and $1.5$ but one should keep in mind that this range depends on the value of the twin lepton Yukawa couplings.

\section{Twin Higgs and lepton superpartners\label{sec:SUSY}}

The Twin Higgs model is not UV complete, and in principle should be treated as an effective field theory limit of some high-energetic theory which solves the big hierarchy problem by introducing colored top partners. In the present work, we focus on SUSY TH models. We do not specify the mechanism that generates the SU(4) invariant quartic term $\lambda$ which can originate either from an $F$-term of a heavy singlet~\cite{Falkowski:2006qq,Chang:2006ra,Craig:2013fga} or a $D$-term of a new gauge symmetry~\cite{badziak2017S,badziak2017M,badziak2017A} and keep $\lambda$ as a free parameter. In this effective description, the particle content of the model is that of MSSM and twin MSSM. 

If the superpartners are light enough, their existence can affect the scalar field thermodynamics at TH energy scales. In particular, the order and strength of phase transitions could be altered. Solution to the hierarchy problem in SUSY TH models generically require the soft mass of the MSSM sparticles and their corresponding twins to preserve the mirror $\mathbb{Z}_2$ symmetry. The LHC set strong lower bounds on masses of colored MSSM states, of about 1 to 2 TeV (with the exact limit depending on the MSSM mass spectrum)~\cite{particle2022}. On the other hand, the LHC limits on slepton masses are much weaker and their masses $\mathcal{O}(100)$~GeV can be still consistent with the experimental data. For such light sleptons even $\mathcal{O}(1)$ twin lepton Yukawa couplings, which allow for strong FOPT as shown in the previous section, do not lead to excessive fine-tuning of the EW scale. 

\subsection{Setup}

In this section, we investigate whether light (twin) sleptons can have non-negligible impact on phase transitions in TH models. We perform this analysis in a simplified model in which the rest of SUSY particles are much heavier and decouple before the onset of phase transitions. We also assume that MSSM-like Higgs bosons and their twin counterparts are decoupled, so during phase transitions only the SM-like and twin Higgs bosons are dynamical. In terms of the MSSM Higgs fields, they are defined as
\begin{equation}
    h_I=H^u_I \sin\beta_I + H^d_I \cos\beta_I \,,
\end{equation}
where up-type Higgs doublet $H^u_I$ couples to up-type fermions while down-type Higgs doublet $H^d_I$ couples to down-type fermions and $ \beta_{I}\equiv\arctan\left(\langle H^u_I\rangle/\langle H^d_I\rangle\right)$. 

Since the Yukawa couplings for the SM leptons are very small, ordinary sleptons do not affect the phase dynamics. However, twin sleptons may have non-negligible impact if twin lepton Yukawa couplings are large enough because they modify the effective potential \eqref{eq:V_tot} via Coleman-Weinberg and thermal corrections. Here, we provide the formulae for slepton masses in order to qualitatively analyze their effect. The tree-level diagonal entries of their mass matrix read
\begin{align}\label{eq:m_sl}
    &\hat{m}^2_{\text{sl}\;R}=\hat{\mu}^2_R+\frac{1}{2}\tilde{y}_{l}^2  h_B^2\cos^2{\beta_{B}}-\frac{1}{4} g'^2 h_B^2 \cos (2\beta_B),\\
    &\hat{m}^2_{\text{sl}\;L}=\hat{\mu}^2_L+\frac{1}{2}\tilde{y}_{l}^2  h_B^2\cos^2{\beta_{B}}-\frac{1}{8} (g^2-\,g'^2) h_B^2 \cos (2\beta_B),
\end{align}
where $\hat{\mu}^2_{L(R)}$ is the soft SUSY breaking mass of the left-handed (right-handed) slepton twin, $\tilde{y}_l^2\equiv\hat{y}_l^2(\tan^2 \beta+1)$ is the supersymmetric twin lepton Yukawa. Mass terms proportional to gauge couplings stem from the D-terms in the SUSY Lagrangian. They were derived from the general expression~\cite{Martin:1997ns} , valid for all squarks and sleptons
\begin{equation}\label{eq:D-term}
    m_{\phi \text{ D-term}}^2=\frac{1}{4}\left(T_{3\phi}-Q_{\phi} \sin^2\theta_W\right)(g^2+g'^2)v_I^2 \cos(2\beta_I).
\end{equation}
Here, $T_{3\phi}$ stands for the third component of the weak isospin while $Q_{\phi}$ corresponds to the electric charge of the left-handed supermultiplet to which the field $\phi$ belongs. 

In general, one can also consider the non-zero mixing between sleptons. In such a case, the most universal, thermally corrected mass matrix reads
\begin{equation}
M=\begin{pmatrix}
\hat{m}^2_{\text{sl}\;R}+\hat{\Pi}_{\text{sl}\;R} &\;\; \frac{1}{\sqrt{2}}\tilde{y}_{l}h_B M_{\text{mix}}\\[5pt]
\frac{1}{\sqrt{2}}\tilde{y}_{l}h_B M_{\text{mix}} &\;\; \hat{m}^2_{\text{sl}\;L}+\hat{\Pi}_{\text{sl}\;L}
\end{pmatrix}.
\end{equation}
Here, $M_{\text{mix}}$ is the mixing parameter which is the combination of MSSM free parameters $M_{\text{Mix}}=A_l\cos(\beta_I)/\tilde{y}_l-\mu \sin(\beta)$, $A_l$ is the soft trilinear lepton coupling while $\mu$ is the supersymmetric Higgsino mass parameter. $\hat{\Pi}_{\text{sl}}(T)$ denotes thermal corrections which have to be included in order to account for the leading order infrared divergences.
We neglect, for simplicity, the mixing between sleptons which is justified unless $\mu$ or $\tan\beta$ are large.

\subsection{Phenomenological impact of sleptons}\label{sec:SUSYphen}

The general analysis of the scalar dynamics in TH augmented by light twin sleptons is a rather complex task. In principle, there are fourteen new free parameters (six soft slepton masses, three mixing parameters, three twin lepton Yukawas, and two SUSY $\beta$ angles) whose values could affect the evolution of Higgs phases. Luckily, an overview of possible slepton effects does not require scanning over all of them. The dynamics of Higgs VEVs depends in the similar way on large groups of the new parameters, so there is no point in considering them separately. Furthermore, it appears that some parameters have negligible overall effects on the considered processes. 

The motivation for introducing light twin sleptons was to obtain strong FOPT in the most promising region of TH parameter space, where fine-tuning is minimal. Hence, till the end of this section we will stick to the benchmark point $\lambda=1$, $\rho=0$ and $f_0=4\,v_A$. As one could see from the previous scans over TH parameter space, the region where FOPT is strong is usually wide and does not contain any holes. Thus, if strong FOPT would take place in the natural region of the model, there is virtually no chance that it will be overlooked.

While there are no experimental constraints on the absolute values of $\tan\beta_{\text{SM}}$ and $\tan\beta_{\text{TS}}$ the cancellation of the loop corrections to SM Higgs mass requires them to be roughly the same. For simplicity, throughout the phenomenological analysis we set them equal. It is important to admit that for low $\tan \beta$ values the Landau pole of the top Yukawa (and its twin) would appear at relatively low energy scales. With the help of full one-loop MSSM renormalization group equations (RGEs)~\cite{Martin:1997ns} we have numerically checked that the top Yukawa Landau pole is above $10^4\;(10^{16})\;\text{GeV}$ for $\tan\beta\gtrsim0.4\; (1.5)$. If the top quark is charged under some extra gauge symmetry, as it is the case e.g. in SUSY D-term TH models~\cite{badziak2017S,badziak2017M,badziak2017A}, the top Yukawa coupling may be asymptotically free even for smaller values of $\tan\beta$.

\begin{figure}[t]
    \centering
    \subcaptionbox{The SM transition strength. \label{fig:scanYSMm}}{\includegraphics[scale=0.42]{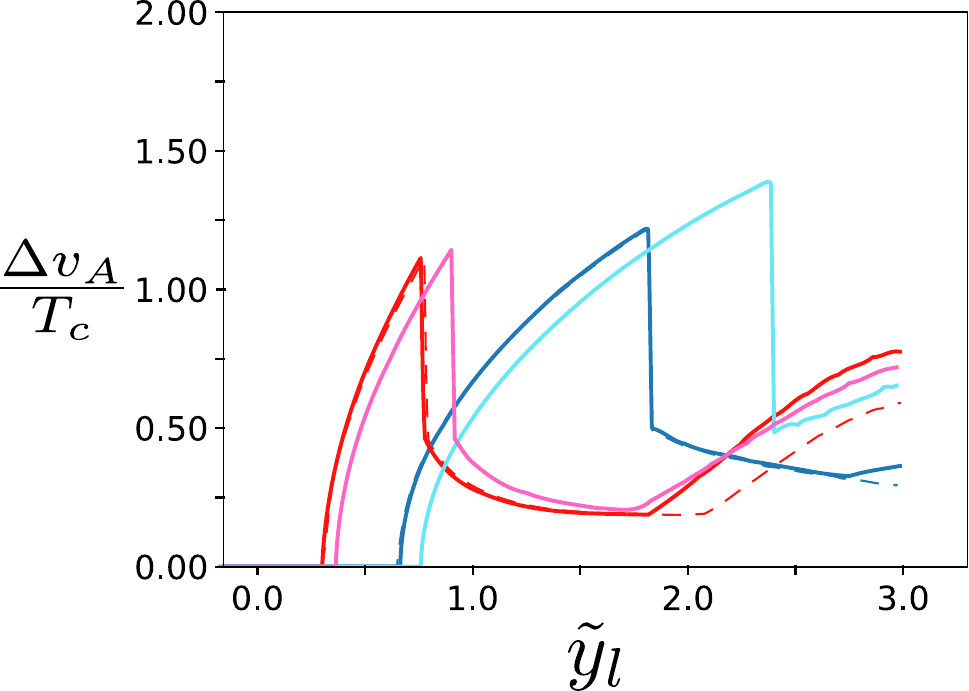}}
    \hspace{5mm}
    \subcaptionbox{The TS transition strength. \label{fig:scanYDSm}}{
    \includegraphics[scale=0.42]{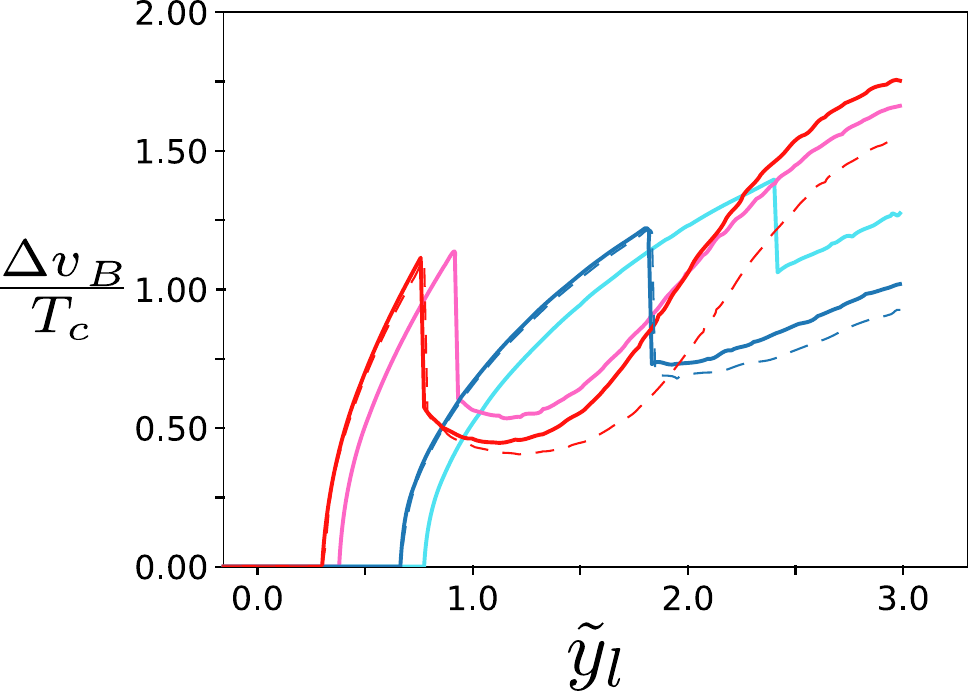}}
    \caption{The phase transition strength $\Delta v_I(T_c)/T_c$ versus twin lepton Yukawa coupling $\tilde{y}_l$ in SUSY TH model with light sleptons for $\tan\beta=2$. The blue and cyan lines correspond to the scenario with enhanced $\tilde{y}_l$ for only one family of twin leptons, while the pink and red lines correspond to Yukawa couplings of all three twin leptons set to $\tilde{y}_l$. 
    The cyan and pink lines denote the results for right-handed sleptons only, wheres the red and blue lines correspond to the scenario with both slepton chiralities. Solid lines denote the results with $\hat{\mu}_{\text{R (L)}}=90\;\text{GeV}$ while dashed lines for those with $\hat{\mu}_{\text{R (L)}}=290\;\text{GeV}$. }
    \label{fig:scanYm}
\end{figure}

In figure~\ref{fig:scanYm} we demonstrate how sleptons affect scalar phase transition for several representative examples. Here, we set $\tan\beta=2$ and assume $\mathbb{Z}_2$ symmetry between the soft masses in the model sectors. We revisit these assumptions later in this section.

We start with the simple TH variant where twin tau Yukawa is enhanced and the right-handed stau with its twin are the only SUSY partners present at thermal equilibrium. The sleptons soft masses are set to $90\;\text{GeV}$ to keep the physical mass of stau above the LEP bound~\cite{particle2022,ALEPH:2001oot,DELPHI:2003uqw,OPAL:2003nhx,L3:2003fyi}. The LHC limits on the right-handed stau are still typically weaker than those from LEP~\cite{CMS:2022rqk}.

We also consider the case where all twin lepton Yukawas are large and all right-handed sleptons are in thermal equilibrium. For simplicity, we set all the twin lepton Yukawas equal. This assumption should not significantly affect phase dynamics, as all twin sleptons contribute to the effective potential in exactly the same way. The LEP lower bounds for all slepton masses are very close to those for stau~\cite{particle2022,ALEPH:2001oot,DELPHI:2003uqw,OPAL:2003nhx,L3:2003fyi} so all soft masses were set to $90\;\text{GeV}$. The LHC limits on the masses of the right-handed smuon and selectron are stronger than those for the right-handed stau but can be easily evaded if the mass splitting between sleptons and the lightest supersymmetric particle (LSP) is smaller than about 50 GeV~\cite{ATLAS:2019lff}.

 We analyze scenarios with both sleptons chiralities (but no mixing). The LHC limits for degenerate left-handed and right-handed sleptons are much stronger than the corresponding limits on the right-handed sleptons so in this scenario sleptons with masses of about 100 GeV are allowed only for fine-tuned values of the mass splitting between sleptons and the LSP~\cite{ATLAS:2019lng,ATLAS:2022fxl}. Thus, we present results not only for $\hat{\mu}_{\text{sl}}\equiv\hat{\mu}_{\,\text{L}}=\hat{\mu}_{\text{R}}=90\;\text{GeV}$ but also for $\hat{\mu}_{\text{sl}}=290\;\text{GeV}$. In the latter case, the physical masses of sleptons are around $300\;\text{GeV}$ so the LHC bounds are satisfied as long as the mass splitting between sleptons and the LSP is below about 100 GeV~\cite{ATLAS:2019lff}. 

\begin{figure}[t]
    \centering
    \subcaptionbox{The SM transition strength. \label{fig:scanYBetaSM}}{
    \includegraphics[scale=0.5]{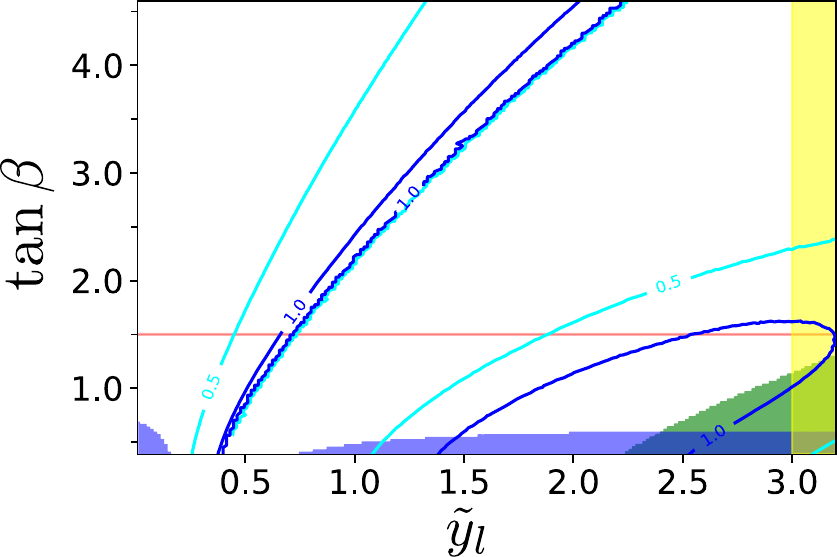}}
    \hspace{10pt}
    \subcaptionbox{The TS transition strength.\label{fig:scanYBetaDS}}{
    \includegraphics[scale=0.5]{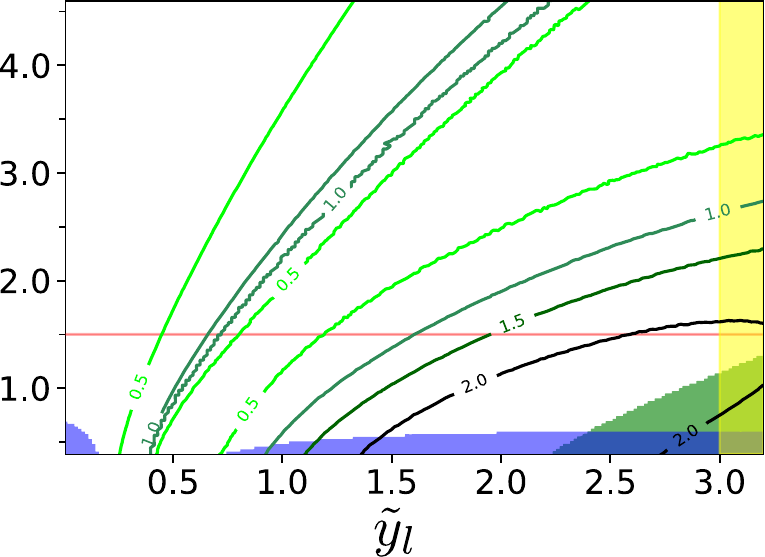}}
    \caption{Contour plots of the phase transition strength $\Delta v_I(T_c)/T_c$ in the $\tilde{y}_l$-$\tan\beta$ plane in SUSY TH model with light right- and left-handed sleptons from all three families. The soft slepton masses were set to $\mu_{\text{sl}}=\hat{\mu}_{\text{sl}}=100\;\text{GeV}$. In accordance with MSSM RGEs the Landau pole of the top Yukawa appears below $10^{16}\;\text{GeV}$ if $\tan\beta\lesssim1.5$ (region under the red line). We cut the plot at $\tan\beta=0.4$ since for smaller $\tan\beta$ values the Landau pole would appear below $10^{4}\;\text{GeV}$. In the yellow region a Landau pole for the twin lepton Yukawa coupling appears below $10\,f_0$. In the blue region the MSSM slepton mass squared was found negative at some point of the thermal evolution.  In the green region sphaleron wash-out of SM baryon asymmetry occurs only at $T>T_c$.}
    \label{fig:scanYBeta}
\end{figure}

It can be seen in figure~\ref{fig:scanYm}, that strong FOPT satisfying the condition \eqref{eq:CondvT1} can be met in all scenarios for the moderate $\tilde{y}_l$ values. In fact, the results of the scan presented in figure~\ref{fig:scanYm} resemble the results in section~\ref{sec:lept} (keep in mind that $\tilde{y}_l=\hat{y}_{l} /\cos\beta$) although this similarity is not exact due to slepton effects, D-terms and thermal mass explicit dependence on $\tan \beta$. As one could expect, the transition type is not affected when twin Yukawas are small because in this regime twin leptons and their superpartners contribution to the effective potential is negligible. Only when $\tilde{y}_l$ exceeds some critical value (which depends on the scenario) its influence on the transition type becomes important. Then, for larger $\tilde{y}_l$ values a temporal drop in $\Delta v_I (T_c)/T_c$ occurs. It is caused by the new intermediate phase that appears between the minima previously separated by the potential barrier. 

The sleptonic impact on phase transition order and strength was observed in all considered scenarios. For very large twin lepton Yukawa couplings $\tilde{y}_l\gtrsim 2$ which are close to the boundary of validity of perturbative calculations, we observe strong FOPT induced by light twin sleptons. Avoiding a Landau pole for the twin lepton Yukawa coupling below $10\,f_0$ sets an upper bound on $\tilde{y}_l\lesssim3$. We admit, that for the allowed values of $\tilde{y}_l$ only in the scenario with three families of light sleptons with large twin Yukawa couplings the phase transition induced by twin sleptons can be stronger than that induced by twin leptons.

Let us now discuss a dependence of the strength of the phase transition on $\tan\beta$. For large $\tan\beta$ the SM-like Higgs boson and its twin are strongly dominated by the corresponding up-type Higgs fields $H^u_I$. On the other hand, twin sleptons couple to the down-type Higgs field $H^d_B$. Therefore, the impact of twin sleptons on the strength transition decreases with $\tan\beta$. This can be seen in figure~\ref{fig:scanYBeta} where we plot contours of $\Delta v_I/T_c$ in the $\tilde{y}_l$-$\tan\beta$ plane for a benchmark scenario with all three families of the right- and left-handed twin sleptons with the soft slepton masses fixed to $100\;\text{GeV}$~\footnote{We choose this scenario as it offers the strongest transition in the TS and thus gives best prospects for DS baryogenesis and emission of the detectable wave gravitational background (see section \ref{sec:GWbackground} for details)}. For small $\tan\beta$ the phase transition can become very strong with $v_B/T_c\gtrsim2$ for $\tan\beta\lesssim1.5$ and perturbative values of twin lepton Yukawa couplings. For example, for $\tan\beta\approx1$ slepton-induced FOPT with $v_B/T_c>1$ requires $\tilde{y}_l\gtrsim1$ (in addition to the one induced by twin leptons which occurs for $\tilde{y}_l\approx0.5$). These coupling values correspond to a region in parameter space in which the top and twin tau Yukawa couplings are perturbative up to scales many orders of magnitude above the EW scale.

We observe that when $\tan\beta<1$ the D-term contributions in the expression for slepton masses \eqref{eq:m_sl} become negative. In the SM sector lepton Yukawas are tiny and negative D-terms could dominate over other terms including thermal mass, provided $h_A(T)$ is big. This is indeed the case in the blue region in figure~\ref{fig:scanYBeta}, where the physical masses of SM sleptons were found negative at some $T>0$, which indicates spontaneous breaking of electromagnetic symmetry.

\begin{figure}[t]
    \centering
    \includegraphics[scale=0.44]{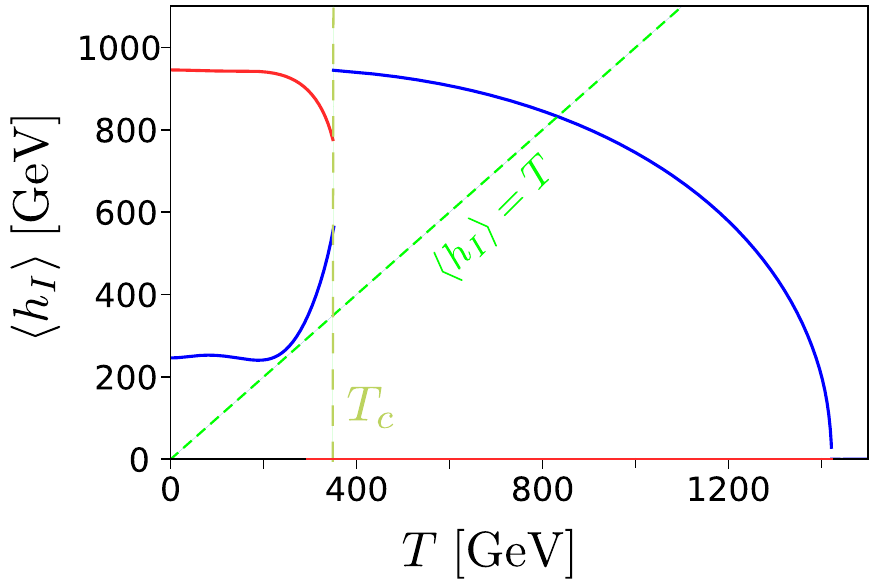}
    \hspace{5mm}
    \includegraphics[scale=0.44]{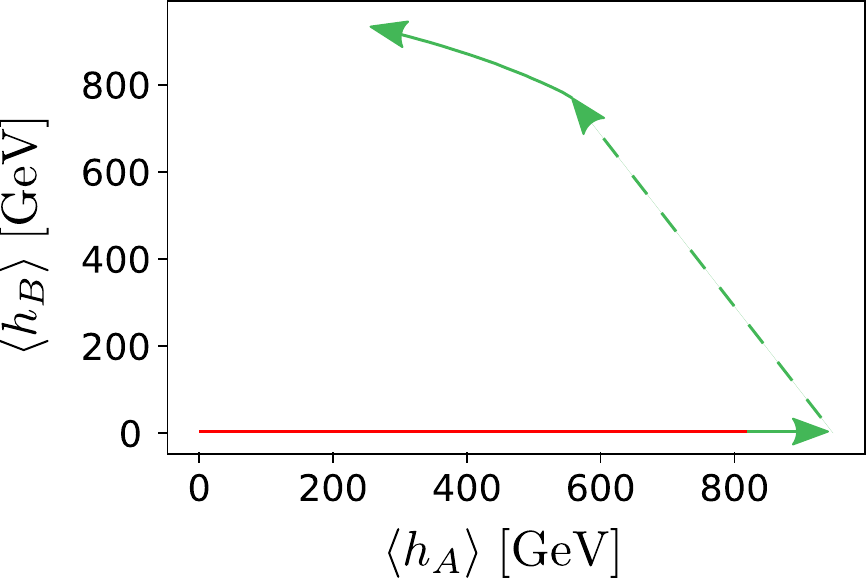}
    \hspace{10pt}
    \caption{Left plot: $h_A(T)$ (blue line) and $h_B(T)$ (red line) VEVs during thermal evolution. Right plot: path in the $\langle h_A\rangle-\langle h_B\rangle$ space with the part of curve where SM sphalerons are suppressed (i.e. $v_{A}(T)/T>1$) marked in green. The plots were completed at the test point $\tilde{y}_l=2.7$, $\tan\beta=0.8$, $\mu_{\text{sl}}=\hat{\mu}_{\text{sl}}=100$~GeV placed in the green region in figure~\ref{fig:scanYBeta}. Baryon asymmetry wash-out does not occur below $T\approx800\; \text{GeV}$.}
    \label{fig:pathlowb}
\end{figure}

We also found that for large (but still perturbative) values of twin lepton Yukawa coupling and $\tan\beta\lesssim1$ the thermal evolution of the fields is qualitatively different. This region is depicted in green in figure~\ref{fig:scanYBeta} and the transition path for a representative point is displayed in figure~\ref{fig:pathlowb}. In contrast to transition paths for larger $\tan\beta$, after the SM-like Higgs gets a VEV at temperature around $f_0$, the EW symmetry is never restored for lower temperatures. As a consequence, SM sphalerons are active only at temperatures significantly above the EW scale. For the benchmark point in figure~\ref{fig:pathlowb} the SM sphaleron decoupling occurs around 800~GeV, so there is no wash-out of the SM baryon asymmetry below that temperature. Such cosmological evolution may have important implications for models of baryogenesis.

\begin{figure}[t]
    \centering
   \includegraphics[scale=0.43]{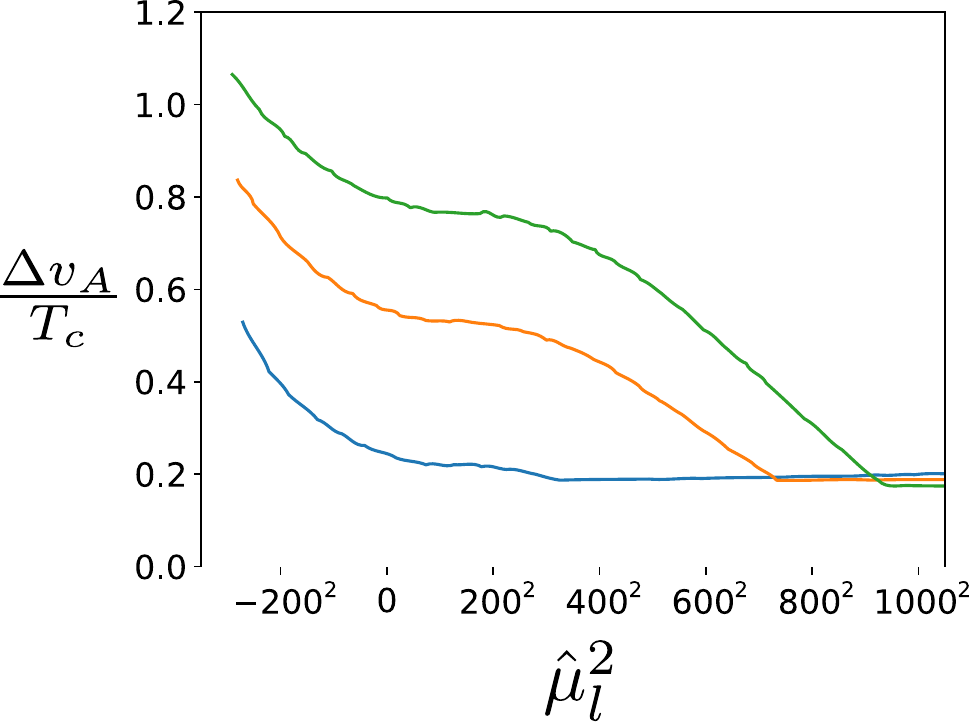}
    \hspace{5mm}
    \includegraphics[scale=0.43]{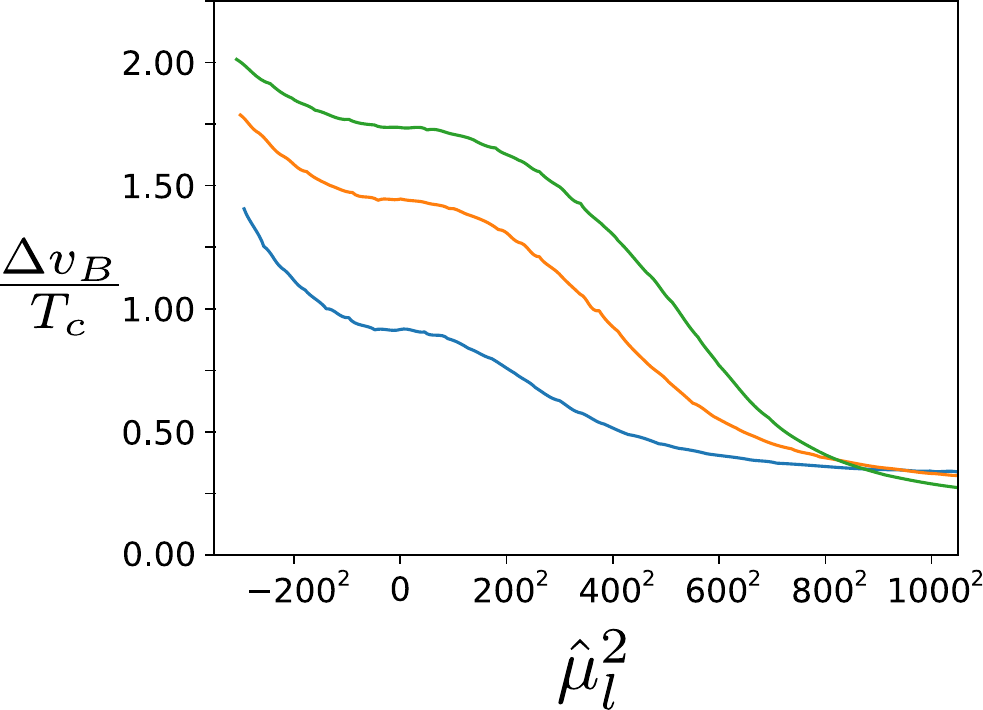}
    \caption{The phase transition strength $\Delta v_I(T_c)/T_c$ vs twin slepton soft mass squared $\hat{\mu}^2_{\text{sl}}$ (equal among all the three generations of the right- and left-handed twin sleptons) for $\tan\beta=2$ and $\mu_{\text{sl}}=90$~GeV. Blue, orange and green lines correspond to $\tilde{y}_l$ set  to $2$, $2.5,$ and $3$, respectively. Each line ends at the point where one of the physical twin slepton masses was found negative.}
    \label{fig:scanLmu}
\end{figure}

Let us now discuss how the strength of the phase transition depends on the value of the soft twin slepton masses. In figure~\ref{fig:scanLmu} we present $\Delta v_I(T_c)/T_c$ as a function of $\hat{\mu}_{\text{sl}}^2$ being a common soft mass for all three families of the right- and left-handed twin sleptons. We fix $\tan\beta=2$ and show the results for three choices of twin lepton Yukawa couplings (equal among the generations): $2$, $2.5$ and $3$. We see that the phase transition strength $\Delta v_B(T_c)/T_c$ is anti-correlated with $\hat{\mu}_{\text{sl}}^2$, leading to stronger transitions for smaller soft twin slepton mass squared. On the other hand, when $\hat{\mu}_{\text{sl}}^2 \gtrsim f_0^2$ twin sleptons effectively decouple, and we recover results obtained for non-supersymmetric TH with enhanced twin lepton Yukawas.

In figure~\ref{fig:scanLmu} we also see that $v_B(T_c)/T_c$ grows significantly when $\hat{\mu}_{\text{sl}}^2$ becomes negative. For large range of negative values of $\hat{\mu}_{\text{sl}}^2$ the physical masses squared of twin sleptons remain positive due to large positive supersymmetric contributions enhanced by large values of $v_B$ and $\tilde{y}_l$. However, large negative values of $\mu_{\text{sl}}^2$ make the MSSM sleptons tachyonic. Hence, enhancing the strength of twin phase transition by negative values of $\hat{\mu}_{\text{sl}}^2$ requires $\mathbb{Z}_2$-breaking in the soft slepton masses. Such $\mathbb{Z}_2$-breaking generically occurs for large twin lepton Yukawa coupling due to negative one-loop corrections $\sim \tilde{y}_l^2$ to twin slepton soft masses, while the corresponding corrections to MSSM sleptons are negligible. Thus, the slepton soft masses consist of two parts: a positive $\mathbb{Z}_2$-preserving contribution and the $\mathbb{Z}_2$-breaking negative loop corrections which, for sufficiently large $\tilde{y}_l$, may become dominant. In figure~\ref{fig:scanLmu} we set $\mu_{\text{sl}}=90$~GeV in order to satisfy the experimental bounds on slepton masses but $\Delta v_B(T_c)/T_c$ does not depend much on a specific value of $\mu_{\text{sl}}$.  \newpage

 In summary, the sphaleron decoupling condition \eqref{eq:CondvT1} in the twin sector is satisfied in all considered scenarios, at least for some values of $\tilde{y}_l$. Even though the SM sphalerons never decouple in a way that could lead to the EW-like baryogenesis in the SM sector, there is a region of parameter space where baryon asymmetry wash-out does not occur. This opens a way for baryon asymmetry transfer from the twin sector and/or inheritance of baryon asymmetry produced in the earlier stages of the evolution of the Universe.

\section{Stochastic gravitational wave background\label{sec:GWbackground}}

 The strong FOPTs are often followed by emission of the stochastic wave gravitational background. Since there are regions of the model parameter space where such transitions are feasible, one may numerically simulate corresponding GW spectra. If, for some parameter choice, the obtained signal falls in the range of any existing or planned detector, one will obtain experimental predictions that could be verified in the near future.

The accurate estimation of GW spectrum emitted during FOPT is a complex problem. Here, we use the standard approximation~\cite{Caprini:2015zlo,Caprini:2019egz}, which is based on four input parameters: the temperature of bubble percolation $T_p$ at which the probability of finding a point still in the false vacuum is 0.7, latent heat released during phase transition $\alpha$, its inverse duration time in the Hubble units $\beta_{\ast}/H_{\ast}$, and the terminal velocity of the new phase bubble walls $v_w$. Further in this section we approximate $T_p$ with the temperature of critical bubble nucleation $T_n$, which is accurate for $\alpha\lesssim 0.2$~\cite{Wang:2020jrd}. To compute the necessary parameters one has to find the profile of the new phase bubbles that nucleate during FOPT. This is a rather complex numerical problem which was handled with dedicated Python package \verb|CosmoTransitions| with custom modifications~\cite{Wainwright:2011kj, marcopriv}. More details about the computation of the input parameters and method used for estimation of GW signal are given in appendix \ref{app:GravT}.

The ratio $\Delta v_I(T_c)/T_c$ used throughout this paper to estimate phase transition strength does not appear in any expression for the GW spectrum. Nonetheless, detectable wave gravitational background cannot be generated when transition is not first-order.The one-loop computations are inaccurate when $\Delta v_I(T_c)/T_c$ is too low (see the appendix~\ref{app:resummation}) and thus, the transitions that were found to be weak first-order at one loop often turn out to be continuous when more accurate computations are performed. For the above reason, in this section we restrict our attention to the parameter values for which the condition \eqref{eq:CondPert} is fulfilled.

\begin{table}[t]
\centering
\begin{tabular}{|llllllll|}
\hline
\multicolumn{1}{|c|}{\textbf{line style}} & \multicolumn{1}{c|}{$\mathbf{f_0}$} & \multicolumn{1}{c|}{$\boldsymbol{\lambda}$} &  \multicolumn{1}{c|}{$\boldsymbol\kappa$}  & \multicolumn{1}{c|}{$\mathbf{T_n}$} & \multicolumn{1}{c|}{$\boldsymbol{\alpha}$} & \multicolumn{1}{c|}{$\mathbf{\boldsymbol\beta_{\ast}/H_{\ast}}$}  & \multicolumn{1}{c|}{$\boldsymbol{v_w}$}  \\ \hline

\multicolumn{1}{|c|}{solid red}  & \multicolumn{1}{c|}{$4\,v_A$} & \multicolumn{1}{c|}{$1$} & \multicolumn{1}{c|}{$-0.1$}  & \multicolumn{1}{c|}{$1172$} & \multicolumn{1}{c|}{$1.3\times 10^{-4}$} & \multicolumn{1}{c|}{$5.1\times10^3$}  & \multicolumn{1}{c|}{$0.39$} \\ \hline

\multicolumn{1}{|c|}{dashed red}  & \multicolumn{1}{c|}{$4\,v_A$} &  \multicolumn{1}{c|}{$1$} & \multicolumn{1}{c|}{$-0.25$} & \multicolumn{1}{c|}{$821$}  & \multicolumn{1}{c|}{$9.4\times10^{-4}$} & \multicolumn{1}{c|}{$1.5\times10^4$} & \multicolumn{1}{c|}{$0.33$} \\ \hline

\multicolumn{1}{|c|}{solid blue}  & \multicolumn{1}{c|}{$8\,v_A$} &  \multicolumn{1}{c|}{$1$} & \multicolumn{1}{c|}{$-0.1$}  & \multicolumn{1}{c|}{$2171$} & \multicolumn{1}{c|}{$ 1.9\times 10^{-4}$} & \multicolumn{1}{c|}{$2.4\times10^3$}  & \multicolumn{1}{c|}{$0.51$} \\ \hline

\multicolumn{1}{|c|}{dashed blue}  & \multicolumn{1}{c|}{$8\,v_A$} &  \multicolumn{1}{c|}{$1$} & \multicolumn{1}{c|}{$-0.25$} & \multicolumn{1}{c|}{$1422$} & \multicolumn{1}{c|}{$1.6\times 10^{-3}$} & \multicolumn{1}{c|}{$4.8\times10^3$}  & \multicolumn{1}{c|}{$0.54$} \\ \hline
\end{tabular}
\caption{Test points for the UV agnostic TH. The corresponding GW spectra are plotted in figure \ref{fig:GWspectraTH}.}
\label{tab:benchmark3}
\end{table}

\begin{figure}[t]
    \centering
    \subcaptionbox{GW spectra in UV agnostic TH model.\label{fig:GWspectraTH}}{\includegraphics[scale=0.40]{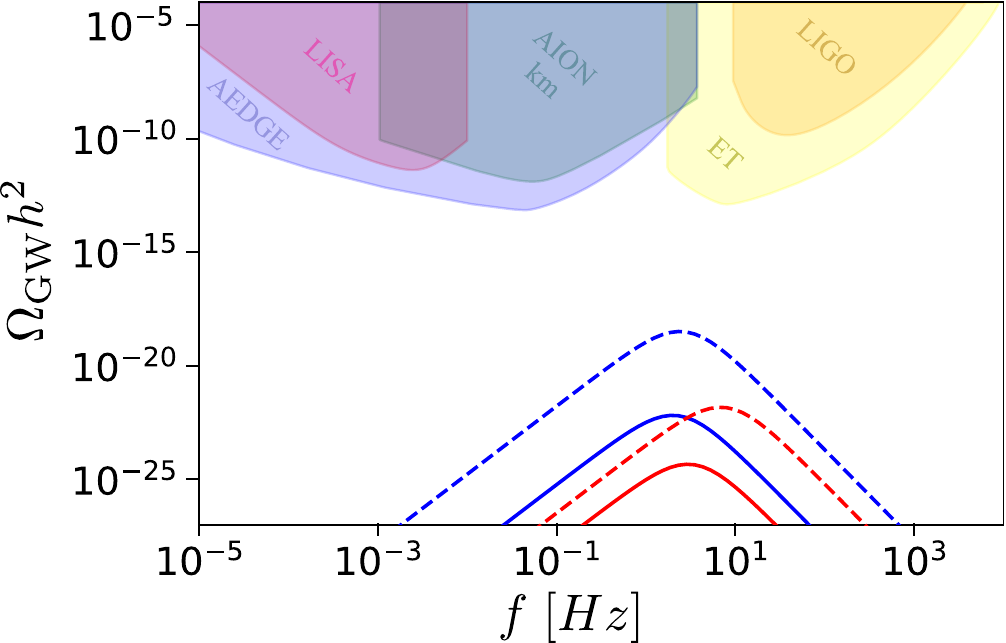}}
    \hspace{5mm}
    \subcaptionbox{GW spectra in SUSY TH model.\label{fig:GWspectraSUSY}}{
    \includegraphics[scale=0.40]{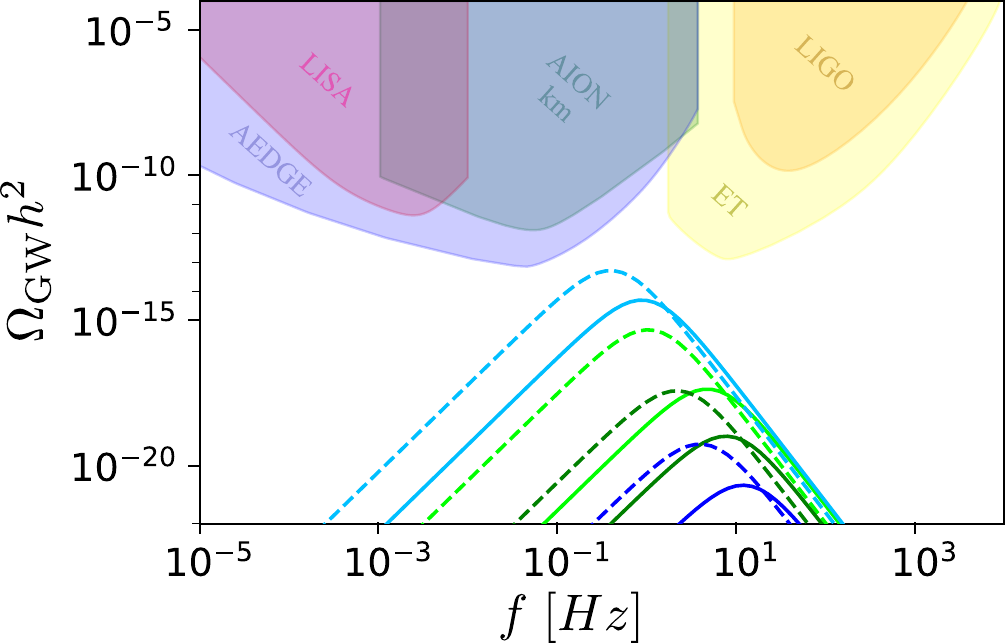}}
    \caption{GW spectra for the benchmark points presented in tables~\ref{tab:benchmark3} and \ref{tab:benchmark4}. Shaded regions mark the integrated power-law sensitivity~\cite{Thrane:2013oya} of the existing and future GW detectors: LIGO~\cite{collaboration2015advanced}, LISA~\cite{Bartolo:2016ami,Caprini:2019pxz}, AION~\cite{Badurina:2019hst}, AEDGE~\cite{Bertoldi:2021rqk} and Einstein Telescope~\cite{Punturo:2010zz,Hild:2010id}. The LISA sensitivity was plotted based on data from the web tool \texttt{PTPlot}~\cite{Caprini:2019egz}. Data about the rest of detector ranges courtesy of M. Lewicki~\cite{lewickipriv}.}
    \label{fig:GWspectra}
\end{figure}

In order to accurately identify regions in the parameter space where potentially detectable gravitational wave background is generated, one should run full scan over the parameter values for which FOPT occurs. Here, we refrained from performing such scan, but instead computed the GW signal for carefully chosen test points, listed in tables~\ref{tab:benchmark3} and \ref{tab:benchmark4}. For the UV agnostic TH we use the free parameter values for which the strong (in terms of $\Delta v_I(T_c)/T_c$) FOPT happens but no significant tuning is present\footnote{We confirmed numerically that for the UV agnostic TH latent heat of the transition $\alpha$ and hence also the GW signal is positively correlated with the ratio $\Delta v_I(T_c)/T_c$.}. In the case of SUSY extension, we restrict our attention to benchmark points with $\lambda=1$, $\rho=0$, $f_0=4\,v_A$ and consider different values of twin lepton Yukawas, slepton soft masses, and $\tan\beta$ values for two distinct scenarios introduced in section \ref{sec:SUSYphen}. The corresponding spectra together with present and future detector sensitivities are shown in figure~\ref{fig:GWspectra}.

\begin{table}[t]
\centering
\begin{tabular}{|llllllll|}
\hline

\multicolumn{1}{|c|}{\textbf{line style}}  &  \multicolumn{1}{c|}{$\mathbf{\hat{n}_{l}}$}  & \multicolumn{1}{c|}{\textbf{tan}$\boldsymbol\beta$} & \multicolumn{1}{c|}{$\mathbf{\hat{\boldsymbol\mu}^2_{\text{sl}}}$ \textbf{[GeV}${}^2$\textbf{]}}   & \multicolumn{1}{c|}{$\mathbf{T_n}$ \textbf{[GeV]}} & \multicolumn{1}{c|}{$\boldsymbol \alpha$} & \multicolumn{1}{c|}{$\mathbf{\boldsymbol\beta_{\ast}/H_{\ast}}$}  & \multicolumn{1}{c|}{$\mathbf{v_w}$}    \\ \hline      

\multicolumn{1}{|c|}{solid green}  & \multicolumn{1}{c|}{$2$} & \multicolumn{1}{c|}{$2$} & \multicolumn{1}{c|}{$90^2$}   & \multicolumn{1}{c|}{$586$} & \multicolumn{1}{c|}{$3.3\times10^{-3}$} & \multicolumn{1}{c|}{$2.3\times10^4$}  & \multicolumn{1}{c|}{$0.21$} \\ \hline
\multicolumn{1}{|c|}{dashed green}  & \multicolumn{1}{c|}{$2$} &  \multicolumn{1}{c|}{$2$} & \multicolumn{1}{c|}{$-300^2$}   & \multicolumn{1}{c|}{$565$} & \multicolumn{1}{c|}{$4.3\times10^{-3}$} & \multicolumn{1}{c|}{$1.1\times10^4$}  & \multicolumn{1}{c|}{$0.30$} \\ \hline
\multicolumn{1}{|c|}{solid blue}  & \multicolumn{1}{c|}{$2$} & \multicolumn{1}{c|}{$1$} & \multicolumn{1}{c|}{$90^2$}   & \multicolumn{1}{c|}{$514$} & \multicolumn{1}{c|}{$7.6\times10^{-3}$} & \multicolumn{1}{c|}{$1.7\times10^4$}  & \multicolumn{1}{c|}{$0.23$} \\ \hline
\multicolumn{1}{|c|}{dashed blue}  & \multicolumn{1}{c|}{$2$} & \multicolumn{1}{c|}{$1$} & \multicolumn{1}{c|}{$-375^2$}   & \multicolumn{1}{c|}{$491$} & \multicolumn{1}{c|}{$1.0\times10^{-2}$} & \multicolumn{1}{c|}{$7.9\times10^3$}  & \multicolumn{1}{c|}{$0.34$} \\ \thickhline

\multicolumn{1}{|c|}{solid light green} & \multicolumn{1}{c|}{$12$} &  \multicolumn{1}{c|}{$2$} & \multicolumn{1}{c|}{$90^2$}  & \multicolumn{1}{c|}{$402$} & \multicolumn{1}{c|}{$1.9\times10^{-2}$} & \multicolumn{1}{c|}{$1.5\times10^4$}  & \multicolumn{1}{c|}{$0.24$} \\ \hline
\multicolumn{1}{|c|}{dashed light green} & \multicolumn{1}{c|}{$12$} &  \multicolumn{1}{c|}{$2$} & \multicolumn{1}{c|}{$-310^2$}  & \multicolumn{1}{c|}{$375$} & \multicolumn{1}{c|}{$2.8\times10^{-2}$} & \multicolumn{1}{c|}{$5.4\times10^3$}  & \multicolumn{1}{c|}{$0.38$} \\ \hline
\multicolumn{1}{|c|}{solid light blue}  & \multicolumn{1}{c|}{$12$} &  \multicolumn{1}{c|}{$1$} & \multicolumn{1}{c|}{$90^2$}  & \multicolumn{1}{c|}{$336$}  & \multicolumn{1}{c|}{$5.0\times10^{-2}$} & \multicolumn{1}{c|}{$5.5\times10^3$} & \multicolumn{1}{c|}{$0.40$} \\ \hline
\multicolumn{1}{|c|}{dashed light blue}  & \multicolumn{1}{c|}{$12$} &  \multicolumn{1}{c|}{$1$} & \multicolumn{1}{c|}{$-230^2$}  & \multicolumn{1}{c|}{$317$}  & \multicolumn{1}{c|}{$6.2\times10^{-2}$} & \multicolumn{1}{c|}{$3.2\times10^3$} & \multicolumn{1}{c|}{$0.49$} \\ \hline

\end{tabular}
\caption{Test points for the SUSY TH model with light sleptons and large twin lepton Yukawa couplings. We have considered two scenarios--one with a single twin lepton with enhanced Yukawa and right-handed slepton ($\hat{n}_{\text{sl}}=2$) and another, where all Yukawas of charged lepton twins are big and sleptons with both chiralities are active ($\hat{n}_{\text{sl}}=12$). We fix $\tilde{y}_l=3$ since we have numerically verified that in all considered points the GW signal is strongest for the largest $\tilde{y}_l$. The remaining parameters are fixed at $\lambda=1$, $\rho=0$, $f_0=4\,v_A$. The corresponding GW spectra are plotted in figure~\ref{fig:GWspectraSUSY}.}
\label{tab:benchmark4}
\end{table}

In the non-supersymmetric TH model the GW signal is many orders of magnitude below the sensitivity of future detectors. The GW signal increases somewhat with the $SU(4)$ breaking scale $f_0$. This is because the GW signal grows with the transition temperature which is proportional to $f_0$. Nevertheless, even for $f_0=8v_A$ the GW signal is more than five orders of magnitude smaller than the sensitivity of future detectors.

In the supersymmetric extension of TH introduced in section~\ref{sec:SUSY} the transition could be caused either by twin leptons or their SUSY partners. In the twin lepton-induced case, which corresponds to the first maximum on the plots in figure~\ref{fig:scanYm}, we found the GW signal to be very weak.
Therefore, in our numerical analysis we have focused on the slepton-induced transition, which occurs close to the second maximum on the plots in figure~\ref{fig:scanYm}. We found that the most promising case for generation large GW signal is the one in which all three twin lepton Yukawa couplings are large, twin sleptons from all three families are as light as possible and $\tan\beta$ is small. Even though the corresponding GW signal is much stronger than in non-supersymmetric case, its peak frequency is still more than an order of magnitude off from the reach of AEDGE~\cite{Bertoldi:2021rqk} and Einstein Telescope~\cite{Punturo:2010zz,Hild:2010id}. The obtained power-low integrated  spectra fall in the blind spot of currently planned detectors and for a rather large range of frequencies between $\mathcal{10^{-1}}$ and $\mathcal{10^{1}}$~Hz is roughly two orders of magnitude below their sensitivity. Nevertheless, to exclude with certainty the detection possibility of this scenario more accurate estimation of GW spectrum would be necessary~\cite{Croon:2020cgk}.

\section{Conclusions and Outlook}\label{sec:summary}

We have shown that Twin Higgs model features strong first-order phase transition provided that appropriate source of $\mathbb{Z}_2$ symmetry breaking between twin and ordinary matter sectors is present. Thermal evolution of the Universe follows an unusual path in the $h_A$--$h_B$ plane, as seen in figure~\ref{fig:sigmal0}, since it is the SM-like Higgs which acquires a non-zero expectation value first at temperature $\sim f_0$. As the Universe cools down, the FOPT occurs during which the SM-like and twin Higgs change their field expectation values. The VEV of the SM-like Higgs decreases from a large initial value to a vanishing one, which excludes the possibility of EW-like baryogenesis. On the other hand, the VEV of the twin Higgs increases, leading to twin sphaleron decoupling after the transition completes. This opens a possibility for dark baryogenesis that may induce the observed asymmetry of SM baryons if an appropriate portal between the twin and SM sector is present.

Our results also show that it is crucial to analyze two-field thermal evolution of the Higgs fields to get a complete picture of phase transitions in TH models. The previous study~\cite{Fujikura:2018duw}, performed using one-field approximation, did not find first-order phase transitions.

We have proposed two distinct scenarios of $\mathbb{Z}_2$ symmetry breaking that leads to the FOPT described above. In the first one \textit{positive}, big asymmetric quartic coupling is present in the tree-level scalar potential. This way one can indeed obtain strong FOPT within the original TH model without specifying its UV completion. Strong FOPT in this scenario requires hard $\mathbb{Z}_2$ breaking, which leads to some tuning against negative $\mathbb{Z}_2$-preserving but $SU(4)$-breaking quartic coupling to accommodate the 125 GeV Higgs mass. Such region of the UV agnostic TH parameter space has been often left without proper analysis.

The second possibility is to include explicit $\mathbb{Z}_2$ breaking by increasing Yukawa couplings of twin leptons. Interestingly, $\mathbb{Z}_2$ breaking in Yukawa couplings relaxes also the dark radiation problem in TH models~\cite{Barbieri:2016zxn,Barbieri:2017opf} and makes twin neutralino a natural candidate for thermally produced DM~\cite{Badziak:2019zys}. In this scenario, strong FOPT can be induced by twin leptons without introducing hard $\mathbb{Z}_2$ breaking in the scalar potential if twin lepton Yukawa couplings are around one. 

In order to avoid excessive tuning in the Higgs potential stemming from large twin lepton Yukawa couplings the TH model must be UV completed at relatively low scale. We have achieved it by considering SUSY UV completion with light sleptons. Moreover, we found that light twin sleptons lead to even stronger FOPT provided that $\tan\beta\lesssim2$ and twin lepton Yukawa couplings are above one.

It is worth mentioning that in the SUSY TH model with light sleptons characterized by small $\tan\beta\lesssim1$ and very large (but still perturbative) twin Yukawa couplings the EW symmetry is broken for temperatures up to above 1~TeV, see figure~\ref{fig:pathlowb} for a representative thermal field evolution in such a case. In this scenario, wash-out of the SM baryon asymmetry, which might be created at some earlier stages of the evolution of the Universe, is delayed until $T\sim\text{TeV}$.

An important part of this work was devoted to estimation of the spectrum of stochastic gravitational wave background which could be emitted during the FOPT. We found that in the UV agnostic TH the obtained GW spectra are nowhere near the reach of the present or planned detectors. In contrast, in the supersymmetric TH model with light sleptons the gravitational wave signal from FOPTs is stronger by many orders of magnitude. In a small region of parameter space we found that the GW signal for frequencies within reach of AEDGE~\cite{Bertoldi:2021rqk} and Einstein Telescope~\cite{Punturo:2010zz,Hild:2010id} detectors can be close but still about two orders of magnitude below their expected sensitivity. 

Our findings open several new avenues for research. Strong FOPTs during which twin EW sphalerons decouple may allow for generation of twin baryon asymmetry. It would be interesting to investigate scenarios in which such asymmetry is transferred to the SM sector and leads to the observed abundance of the SM baryons. It would be also interesting to investigate other variants of TH models in which $\mathbb{Z}_2$ breaking leads to strong FOPTs. Finding models in which detectable GWs and/or EW symmetry non-restoration up to temperatures much above the EW scale are more generic phenomena would be particularly exciting.

\section*{Acknowledgments}

 The authors would like to thank Keisuke Harigaya, Marek Lewicki, Marco Merchand, Mariano Quiros, Bogumiła Świeżewska and Mateusz Zych for useful discussions and Kohei Fujikura for useful correspondence. This work was partially supported by the National Science Centre, Poland, under research grant no. 2020/38/E/ST2/00243.

\begin{appendices}
\section{Tree-level Higgs masses and VEVs beyond PNGB approximation.}\label{app:VEVandMass}
The exact expressions for the tree-level masses and VEVs of SM and twin Higgses can be derived analytically~\cite{Ahmed:2017psb}. They read
\begin{equation}
v_A^2=\lambda f_0^2\frac{-\sigma+\kappa(1-\frac{\sigma}{\lambda})}{\lambda\rho+\kappa (2\lambda+\rho+\kappa)},\qquad v_B^2=f^2-v_A^2 \,,
\end{equation}
where
\begin{equation}
f^2=f_0^2\frac{\lambda\rho+\kappa(2\lambda-\sigma)}{\lambda\rho+\kappa (2\lambda+\rho+\kappa))}\,,
\end{equation}
The expressions for the tree-level masses of the SM-like Higgs boson, $m_h$, and its twin counterpart, $m_{h'}$, are
\begin{equation}
\begin{aligned}
&m_{h(h')}^2=\rho v_A^2+ f^2 (\lambda+\kappa)\left(1\mp \sqrt{1-A}\right) \,,\\[2pt]
&A\equiv2\frac{v_A^2}{f^2}\frac{\lambda\rho+\kappa(4\lambda+\rho+2\kappa)}{
(\lambda+\kappa)^2}-\frac{v_A^4}{f^4}\frac{4\lambda\rho+\rho^2+\kappa(8\lambda+4\rho+4\kappa)}{
(\lambda+\kappa)^2} \,,
\end{aligned}
\end{equation}
where the upper (lower) sign in the above equation corresponds to $m_h$ ($m_{h'}$). The approximate expressions \eqref{eq:VEV-s} and \eqref{eq:masses} are recovered by expanding the exact tree-level formulas to the leading order in $v_A^2/f^2\approx v_A^2/f^2_0$ and $\kappa/\lambda,\;\sigma/\lambda,\;\rho/\lambda$.
\section{Finite temperature terms in the effective potential}
\subsection{Low and high temperature expansion of the thermal potential\label{app:expansion}}

To evaluate thermal functions $J_B$ and $J_F$ one can directly compute the integrals used in their definitions \eqref{eq:therm}. Alternatively, it is possible to use the expressions for the low and high temperature expansion of those functions. For $T\gg m$ one can use~\cite{Quiros:1999jp}
\begin{align}
&J_B(\frac{m^2}{T^2})\approx -\frac{\pi^4}{45}+\frac{\pi^2}{12}\frac{m^2}{T^2}-\frac{\pi}{6}\left(\frac{m^2}{T^2}\right)^{\frac{3}{2}}-\frac{m^4}{32T^4}\log\left(\frac{m^2}{a_b\,T^2}\right)+\mathcal{O}\left(\frac{m^6}{T^6}\right)\label{eq:JBapp}\\
&J_F(\frac{m^2}{T^2})\approx\frac{7\pi^4}{360}-\frac{\pi^2}{24}\frac{m^2}{T^2}-\frac{m^4}{32T^4}\log\left(\frac{m^2}{a_f\,T^2}\right)+\mathcal{O}\left(\frac{m^6}{T^6}\right)\label{eq:JFapp}
\end{align}
with $a_b=16\,\pi^2 e^{\frac{3}{2}-2\gamma_\text{E}}$ and $a_f=\pi^2 e^{\frac{3}{2}-2\gamma_\text{E}}$, where $\gamma_\text{E}$ is the Euler's constat. 

For $T\leq m$ the corresponding expansion reads~\cite{Curtin:2016urg}
\begin{equation}\label{eq:JBJFapp}
\begin{aligned}
    &J_B(\frac{m^2}{T^2})=-\sum_{n=1}^{\infty} \frac{1}{n^2}\frac{m^2}{T^2} \text{K}_2(n\,\frac{m}{T})\\
    &J_F(\frac{m^2}{T^2})=-\sum_{n=1}^{\infty} \frac{(-1)^n}{n^2}\frac{m^2}{T^2} \text{K}_2(n\,\frac{m}{T})
\end{aligned}
\end{equation}
where $\text{K}_2(x)$ is the Bessel function of the second kind. The above expression for low temperature expansion is less intuitive but it remains surprisingly accurate even for relatively high temperatures. It also does not require including many terms to get a good precision -- two or three are usually sufficient.

The computation of the approximate forms \eqref{eq:JBapp}-\eqref{eq:JBJFapp} is much faster than direct evaluation of the integrals \eqref{eq:therm}. Since in the whole range $J_B$ and $J_F$ are well-approximated either by low or high temperature expansion, to obtain numerical results we use the series defined above rather than $J_B$ and $J_F$ definition.

\subsection{Resummation of infrared divergences }\label{app:resummation}

At high temperatures the non-abelian gauge theories with massive vectors suffer from infrared divergences. The divergent contributions arise in the vicinity of phase transition, where all particles which acquire mass due to Higgs mechanism become massless. The leading order infrared contribution stems from the so-called daisy diagrams. It scales with the number of massless bosonic loops $n$ as $\sim (\mathcal{g}^2 T^2/m^2)^n$, where $\mathcal{g}$ is the characteristic gauge coupling and $m$ characteristic gauge mass scale of the problem. At the phase transition $m\approx \mathcal{g}\phi\sim \mathcal{g} T$ and the scaling factor becomes roughly $1$.

This issue can be alleviated by the resummation of the leading order infrared divergences. After the resummation, the series of most divergent diagrams is suppressed by an additional factor $\mathcal{g}^2T/m(\phi)\sim \mathcal{g}v_c/T_c$\footnote{In principle, in the suppression factor one should use thermally resummed mass $\overline{m}(\phi)$ defined by the eq. \eqref{eq:mT}. However, most divergent contributions come from transverse modes of the massive vectors which do not receive thermal corrections at leading order, and thus $\overline{m}(\phi)\approx m(\phi)$.}. Hence, the rough criterion for the applicability of one-loop approximation of the effective potential \textit{after} the resummation reads
\begin{equation}\label{eq:pert}
\frac{\phi(T)}{T}>\mathcal{g},
\end{equation}
from which one obtains the condition \eqref{eq:CondPert} for TH. 

Strictly speaking, the results derived within one-loop approximation in gauge theories are reliable only when the transition is sufficiently strong, i.e. $\phi(T_c)/T_c\gg \mathcal{g}$ (otherwise, the computations are unreliable in the vicinity of phase transition). Moreover, the accuracy increases with the transition strength. To some extent, the criterion \eqref{eq:pert} may be weakened by the more accurate resummation techniques~\cite{Curtin:2016urg,Croon:2020cgk,Ekstedt:2022zro,Ekstedt:2022bff} which take into account the sub-leading multi-loop corrections. However, their implementation requires much more effort and usually does not improve the accuracy a lot at the temperature at which the strong FOPT occurs. 

In this paper, we consider two fields dynamics, and hence one field criterion \eqref{eq:CondPert} does not necessarily guarantee the perturbativity at phase transition. Up to our knowledge, the proper criterion for the type of FOPT we found in TH is yet unknown. We expect that if the condition \eqref{eq:CondPert} is satisfied in the theory sector with broken symmetry before and after the transition, our perturbation approximation holds at $T=T_c$.

The resummation of the leading order infrared divergences is equivalent with replacing tree-level bosonic masses in the thermal and Colman-Weinberg parts of the effective potential with the effective masses~\cite{Parwani:1991gq}
\begin{equation}\label{eq:mT}
    m^2_{i}(h_I)\rightarrow\overline{m}_{i}^2\equiv m^2_i(h_I)+\Pi(T).
\end{equation}
In case where mass matrix is non-diagonal, the thermal contributions have to be added \textit{prior} to mass diagonalization. 
For instance, the thermally corrected mass matrix for SM and TS Higgses reads
\begin{equation}
   \overline{M}_{h,\;H}=\begin{pmatrix}
3(\lambda+\kappa+\rho)h_A^2+\lambda h_B^2-f_0^2(\lambda-\sigma) +\Pi_{h_A} & 2\lambda h_A h_B \\
2\lambda h_A h_B & 3(\lambda+\kappa) h_B^2+\lambda h_A^2-f_0^2\lambda +\Pi_{h_B}
\end{pmatrix},
\end{equation}
while the $Z-\gamma$ mass matrix is
\begin{equation}
    \overline{M}_{Z,\;\gamma}=\begin{pmatrix}
g^2 h_A^2+\Pi_Z & g g' h_A^2 \\
g g' h_A^2& g'^2 h_A^2+\Pi_{\gamma}
\end{pmatrix}.
\end{equation}
Below, we list all thermal corrections $\Pi(T)$ which were used in the numerical computations. Since these corrections generally depend on the particle content of the theory, the additional contributions appear for particular UV extensions considered in this paper. For the ordinary TH the relevant contributions are
\begin{equation}
\begin{aligned}
&\phat{\Pi}_W=\frac{11}{6}g{\,}^2T^2,\qquad\phat{\Pi}_Z=\frac{11}{6}g^2T^2,\qquad\phat{\Pi}_{\gamma}=\frac{11}{6}g'{}^2T^2,\\
&\Pi_{h_A}=\Big(\frac{5}{6}\lambda+\frac{1}{2}\kappa+\frac{1}{2}\rho+\frac{1}{16}g'^2+\frac{3}{16}g^2+\frac{1}{4}y_t^2\Big)\,T^2,\\
&\Pi_{h_B}=\Big(\frac{5}{6}\lambda+\frac{1}{2}\kappa+\frac{1}{16}g'{}^2+\frac{3}{16}g^2+\frac{1}{4}\hat{y}_t^2\Big)\, T^2,\\
\end{aligned}
\end{equation}
As in the main text, we use here hat to denote TS particles and couplings. The hats are suppressed only in the case of twin Higgs, $g$ and $g'$ for which they are redundant. We stress that only the longitudinal degrees of freedom of the massive vectors receive thermal contributions at the leading order. The transverse modes receive thermal corrections of order $\mathcal{O}(g^4)$ from non-perturbative effects. Those sub-leading non-perturbative corrections were neglected in our numerical computations.

In the UV agnostic TH model with enhanced $\hat{y}_l$, one has to include leptonic thermal contribution to the twin Higgs mass
\begin{equation}\label{eq:mH_lept}
\Pi_{h_B}=\Big(\frac{5}{6}\lambda+\frac{1}{2}\kappa+\frac{1}{16}g'{}^2+\frac{3}{16}g^2+\frac{1}{4}\hat{y}_t^2+ \frac{\hat{n}_l}{48} \hat{y}_l^2 \Big)T^2,
\end{equation}
where $\hat{n}_{l}$ is the number of degrees of freedom coupled with the enhanced Yukawa $\hat{y}_l$. 

As far as light sleptons are considered, one needs to account for their thermal masses~\cite{Comelli:1996vm}
\begin{equation}
\begin{aligned}
&\hat{\Pi}_{\text{sl}\;R}=\Big\{\frac{1}{12}g'^2\big[4+\frac{1}{2}(\hat{n}_{\text{sl, R}}-\hat{n}_{\text{sl, L}})-\cos(2\beta)\big]+\frac{1}{6}\tilde{y}_{l}^2(\cos^2\beta+\theta_{\tilde{e}_{L}})\Big\}T^2,\\
&\hat{\Pi}_{\text{sl}\;L}=\Big\{\frac{1}{4} g^2+\frac{1}{24} g'^2\big[2+\frac{1}{2}(\hat{n}_{\text{sl, L}}-\hat{n}_{\text{sl, R}})+\cos(2\beta)\big]+\frac{1}{12}\tilde{y}_{l}^2(\cos^2\beta+\theta_{\tilde{e}_{R}})\Big\}T^2.
\end{aligned}
\end{equation}
Here, $\hat{n}_{\text{sl, L}}$ ($\hat{n}_{\text{sl, R}}$) is the number of the left-handed (right-handed) twin sleptonic degrees of freedom coupled to twin Higgs via enhanced Yukawa coupling. $\theta_{\tilde{e}_{I}}$ is unity if $\tilde{e}_{I}$ is present in thermal equilibrium and zero if it is decoupled.
In the MSSM-like extension also the thermal masses of TS Higgs and massive vectors coupled with sleptons are modified~\cite{Comelli:1996vm}
\begin{equation}
\begin{aligned}
&\Pi_{h_B}=\Big[\frac{5}{6}\lambda+\frac{1}{2}\kappa+\frac{1}{16}g'{}^2+\frac{3}{16}g^2+\frac{1}{4}\tilde{y}_t^2+ \frac{\hat{n}_l}{48} \tilde{y}_l^2\cos^2(\beta)+ \frac{\hat{n}_{\text{sl}}}{24} \tilde{y}^2_l \cos^2(\beta) \Big]T^2,\\
&\phat{\Pi}_{W}=\phat{\Pi}_{Z}=\Big(\frac{11}{6}+\frac{1}{12}\phat{n}_{\text{sl, L}}\Big)g^2T^2,\qquad\quad \phat{\Pi}_{\gamma}=\Big(\frac{11}{6}+\frac{1}{12}\phat{n}_{\text{sl, L}}+\frac{1}{6}\phat{n}_{\text{sl, R}}\Big)g'{}^2T^2.
\end{aligned}
\end{equation}
where $\hat{n}_{\text{sl}}$ is the total number of twin sleptonic degrees of freedom coupled to twin Higgs via enhanced Yukawa coupling.

\section{Estimation of GW signal\label{app:GravT}}

\subsection{Input parameters}
In order to obtain the power spectrum of GW, one first needs to compute the percolation temperature $T_p\approx T_n$, latent heat of the transition $\alpha$, (inverse) time duration of the transition in Hubble units $\beta_{\ast}/H_{\ast}$ and terminal wall velocity $v_w$. The first three parameters can be relatively easy obtained from the effective potential \eqref{eq:V_tot} and three-dimensional Euclidean action $S_3$~\cite{Lewicki:2021pgr,LISACosmologyWorkingGroup:2022jok}
\begin{align}
    &\alpha=\frac{30}{\pi^2 T_n^4 g_{\ast}}\left(\Delta V_{\text{eff}}(T_n)-\frac{1}{4}T_n\Delta\left.\frac{\partial V_{\text{eff}}}{\partial T}\right|_{T=T_n}\right),\\
    &\frac{\beta_{\ast}}{H_{\ast}}=T\left.\frac{\partial}{\partial T}\left(\frac{S_3}{T}\right)\right|_{T=T_n},
\end{align}
where $\Delta$ denotes the difference between quantities evaluated in the false and true vacua while $g_{\ast}$ is the number of relativistic degrees of freedom in the model at $T=T_n$. 
The terminal bubble wall velocity was obtained using equilibrium approximation~\cite{Lewicki:2021pgr}
\begin{equation}\label{eq:velocity}
    v_w^2\approx\frac{\Delta V_{\text{eff}}}{\Delta V_{\text{eff}}(T_n)-\frac{1}{4}T_n\Delta\left.\frac{\partial V_{\text{eff}}}{\partial T}\right|_{T=T_n}},
\end{equation}
which turns out to be quite accurate and model independent as long as bubbles remain subsonic. When the expression \eqref{eq:velocity} exceeds the speed of sound $c_s^2=1/3$, one can assume that $v_w=1$.

\subsection{Spectrum}
During FOPT there are three main sources of GW: bubble wall collisions, magnetohydrodynamic turbulence and sound waves produced in bubble collisions. Their contribution to the overall GW signal highly depends on the phase transition details. 

At all chosen test points, the friction exerted by plasma was sufficient to stop the acceleration of bubble walls at equilibrium terminal velocity since the run-away condition~\cite{Espinosa:2010hh, Caprini:2015zlo}
\begin{equation}
    \alpha>\alpha_{\infty}\approx\frac{30}{24\pi^2}\frac{\sum_{\text{bos}}\Delta m_i^2 (T=0)+\frac{1}{2}\sum_{\text{ferm}}\Delta m_i^2 (T=0)}{ g_{\ast} T_{n}^2}
\end{equation}
was never fulfilled. Here, $\Delta$ denotes difference between the mass squared evaluated in the new and the old phase. On the other hand, at all test points equilibrium terminal velocities obtained with \eqref{eq:velocity} were subsonic. Thus, we always stay in the regime where bubble wall collisions are a negligible source, and GW generation is dominated by sound waves~\cite{Ellis:2018mja}. In this case, the approximate analytic expression for the GW spectrum reads~\cite{Caprini:2015zlo}
\begin{equation}
    h^2\Omega_{\text{SW}}=2.65\times10^{-6}\left(\frac{H_{\ast}}{\beta_{\ast}}\right)\left(\frac{\kappa_{\text{SW}}\alpha}{\alpha+1}\right)^2\left(\frac{100}{g_{\ast}}\right)^{\frac{1}{3}}v_w\,S_{\text{SW}}(f).
\end{equation}
Here, $\kappa_{\text{SW}}$ is the fraction of the latent energy released in the transition that was transmitted to sound waves. To find its true value, one would need to solve the equations of state for the scalar phase. In this work, we use approximate formulae for $\kappa_{\text{SW}}$ derived in~\cite{Espinosa:2010hh} which turn out to be accurate for a large class of models with polynomial scalar potentials~\cite{Caprini:2019egz}.
The profile function for the sound waves $S_{\text{SW}}(f)$ is given by~\cite{Caprini:2015zlo}
\begin{equation*}
    S_{\text{SW}}(f)=\left(\frac{f}{f_{\text{SW}}}\right)^3\bigg(\frac{7}{4+3\left(\frac{f}{f_{\text{SW}}}\right)^2}\bigg)^{\frac{7}{2}},
\end{equation*}
where 
\begin{equation*}
    f_{\text{SW}}=1.9\times 10^{-5} \text{Hz}\,\frac{1}{v_w} \left(\frac{\beta_{\ast}}{H_{\ast}}\right)\left(\frac{T_{\ast}}{100\, \text{GeV}}\right)\left(\frac{g_\ast}{100}\right)^{\frac{1}{6}}.
\end{equation*}

\end{appendices}
\medskip
\bibliography{sources}

\end{document}